\documentclass[conference]{IEEEtran}
\makeatletter
\def\ps@headings{\def\@oddhead{\mbox{}\scriptsize\rightmark \hfil}\def\@evenhead{\scriptsize\thepage \hfil \leftmark\mbox{}}\def\@oddfoot{}\def\@evenfoot{}}
\makeatother
\usepackage{graphicx}   \usepackage{balance}    \usepackage{subfigure}
\usepackage{multirow}
\usepackage{pbox}
\usepackage{cite}
\usepackage{float}
\usepackage[table]{xcolor}\usepackage{epstopdf}
\usepackage{amsmath}
\usepackage{hhline}
\usepackage{setspace}
\usepackage{bbding}
\usepackage{comment}
\usepackage{xcolor}
\newcommand{\figscale}{0.728}
\newcommand{\system}{Map++}
\usepackage{tabularx}
\newcolumntype{s}{>{\raggedright\arraybackslash}X}
\usepackage{slashbox}
\usepackage[colorinlistoftodos]{todonotes}
\usepackage[pscoord]{eso-pic}
\IEEEoverridecommandlockouts
\IEEEpubid{\makebox[\columnwidth]{978-1-4799-4657-0/14/\$31.00~\copyright~2014 IEEE \hfill}
\hspace{\columnsep}\makebox[\columnwidth]{ }}
\begin{document}\fontsize{10}{11.58}\rm

\title{Map++: A Crowd-sensing System for Automatic Map Semantics Identification}

\author{\IEEEauthorblockN{Heba Aly}
\IEEEauthorblockA{Dept. of Computer and Sys. Eng.\\
Alexandria University, Egypt\\
Email: heba.aly@alexu.edu.eg}
\and
\IEEEauthorblockN{Anas Basalamah}
\IEEEauthorblockA{Comp. Eng. Dept., KACST GIS Tech. Innov. Ctr.\\
Umm Al-Qura Univ., Makkah, Saudi Arabia\\
Email: ambasalamah@uqu.edu.sa}
\and
\IEEEauthorblockN{Moustafa Youssef}
\IEEEauthorblockA{Wireless Research Center\\
Alexandria Univ. and E-JUST, Egypt\\
Email: moustafa.youssef@ejust.edu.eg}}

\maketitle

\begin{abstract}
Digital maps have become a part of our daily life with a number of commercial and free map services. These services have still a huge potential for enhancement with rich semantic information to support a large class of mapping applications.
In this paper, we present \system{}, a system that leverages standard cell-phone sensors in a crowdsensing approach to automatically enrich digital maps with different road semantics like tunnels, bumps, bridges, footbridges, crosswalks, road capacity, among others. Our analysis shows that cell-phones sensors with humans in vehicles or walking get affected by the different road features, which can be mined to extend the features of both free and commercial mapping services. 
We present the design and implementation of \system{} and evaluate it in a large city. Our evaluation shows that we can detect the different semantics accurately with at most 3\% false positive rate and 6\% false negative rate for both vehicle and pedestrian-based features. Moreover, we show that \system{}  has a small energy footprint on the cell-phones, highlighting its promise as a ubiquitous digital maps enriching service.

\end{abstract}

\IEEEpeerreviewmaketitle

\section{Introduction}

Recently, digital maps have gained great attention due to their high economic and social impact; They are integrated into our everyday lives in different forms such as navigation systems, traffic estimation services, location based services, asset tracking applications, and many more. Realizing the economic value of this technology, several giant companies are producing commercial map services including Google Maps\cite{googlemaps}, Yahoo! Maps\cite{yahoomaps}, and Microsoft's Bing Maps\cite{bingmaps}, as well as free services such as OpenStreetMaps \cite{osm}. These map services attract millions of users daily. In 2013, Google announced that its Google Maps service is accessed by over one billion users every month \cite{googleio}.

Typically, these maps are constructed through satellite images, road surveyors, and/or manual entry by trained personnel\cite{navteqhist}. However, with the dynamic changes and richness of the physical world, it is hard to keep these digital maps up-to-date and capture all the physical world road semantics. To address this issue, commercial map companies started to provide tools, e.g. Google's MapMaker\cite{googlemapmaker} and Nokia's HERE Map Creator\cite{heremapcreator}, for users to manually send feedback about their maps, i.e. crowdsource the map updates. This was even generalized to build entire completely-free editable maps such as OpenStreetMap (OSM) \cite{osm} and WikiMapia\cite{wikimapia}. However, these services require active user participation and are subject to intentional incorrect data entry by malicious users.

\begin{figure*}[!t]
\centering
 \subfigure[Satellite image annotated with semantics identified by \system.]{
      \includegraphics[width=0.31\linewidth]{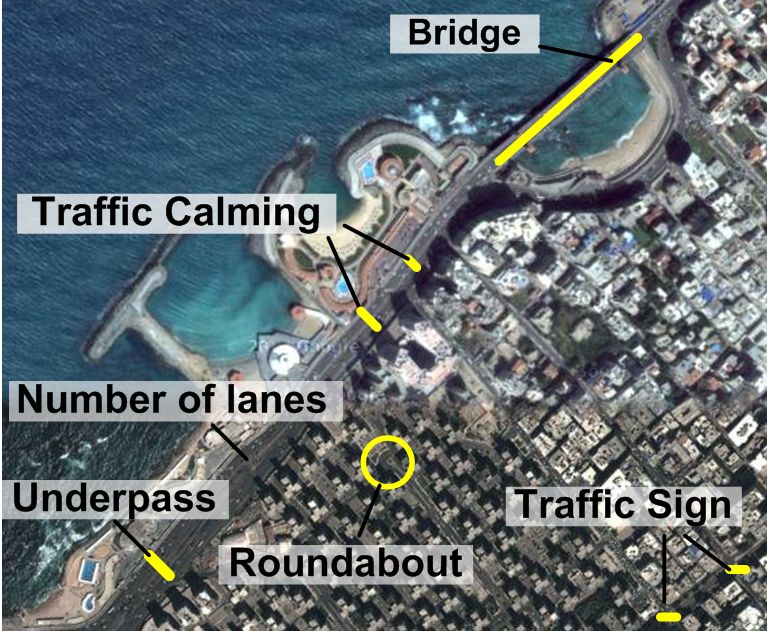}
      \label{fig:satellite}
    }
    \subfigure[Google Maps\cite{googlemaps}]{
      \includegraphics[width=0.31\linewidth]{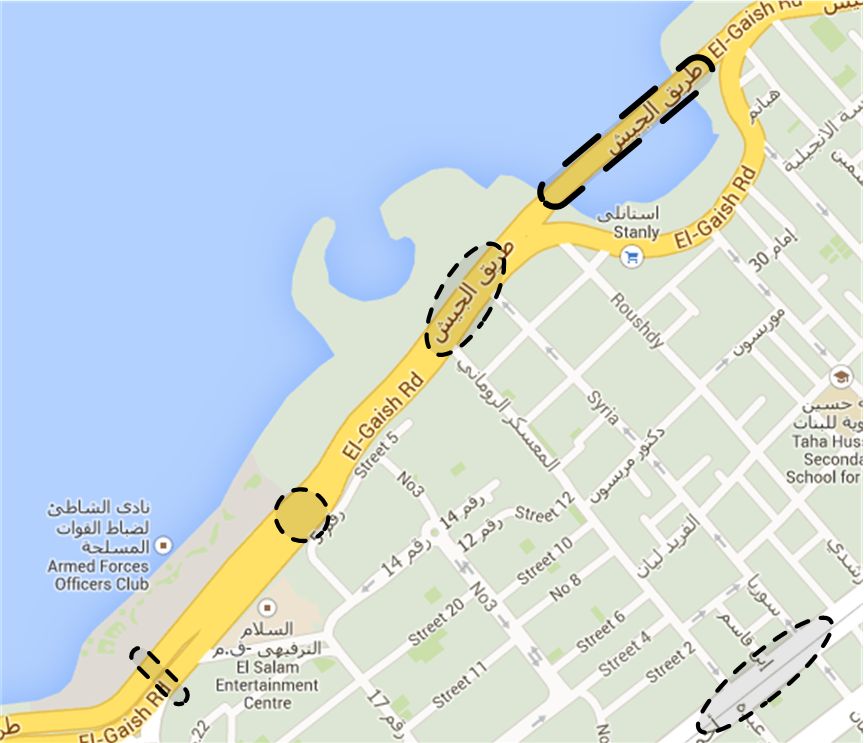}
      \label{fig:google}
    }
    \subfigure[OpenStreetMap\cite{osm}]{
      \includegraphics[width=0.31\linewidth]{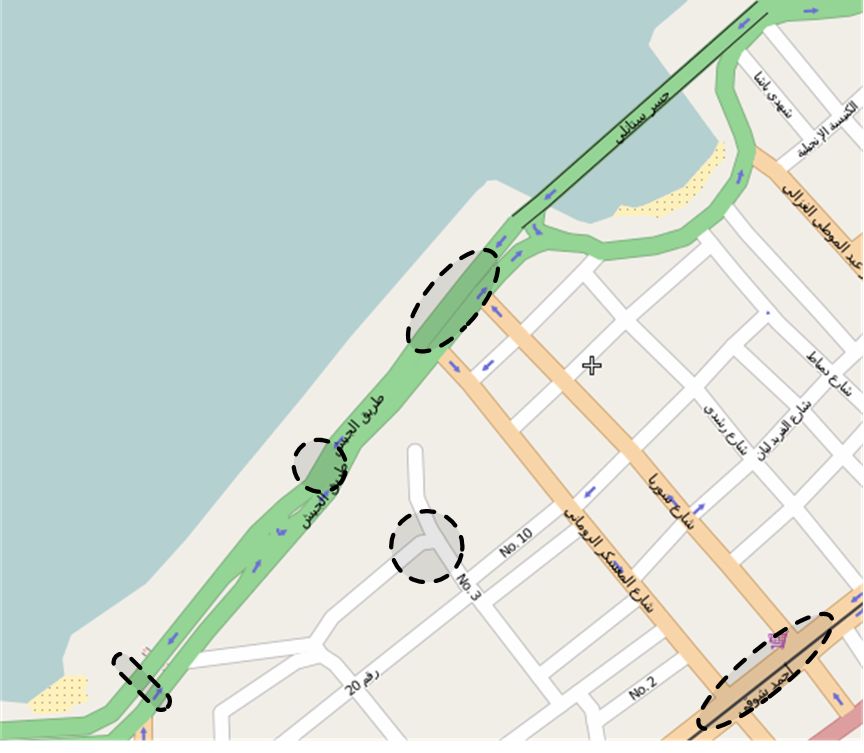}
  \label{fig:osm}
    }
\caption{An example from the city of Alexandria, Egypt showing that both commercial and free mapping services miss a number of semantic features for next generation maps. The satellite image is used to show the actual area and is annotated with the full semantics extracted by \system{}; The majority of these features are missing in all current mapping services ({\it shadowed ellipses}). Google Maps even does not identify the bridge and defines it as a normal road.}
\label{fig:map_comp}
\end{figure*}

With the proliferation of today's sensor-rich mobile devices, cell phones are becoming the bridge between the physical and digital worlds. Researchers leveraged the GPS chips on smart phones to collect traces that can be used  automatically to update existing maps and infer new roads \cite{crowdAtlas,cao2009gps,baier2011mapcorrect}. However, GPS is an energy hungry device and these systems focus only on estimating missing road segments. In summary, all existing mapping services, both commercial and free, miss a large number of semantic features (Figure~\ref{fig:map_comp}) that are a necessity for many of today's map-based applications. For example, navigation systems relay on important semantics to better guide users to their destinations; a short route  may be falsely tempting if traffic lights are hidden from the user, a pedestrian tourist might be deceived when finding out that the road has no sidewalks, city evacuation planning might be ineffective if maps are not tagged with the number of lanes, a driver might be at risk of an accident if his map does not show the road bumps ahead, and a person with disability needs a map that shows the elevator-enabled subway stations.

In this paper, we present \system{} as a system that leverages the ubiquitous sensors available in commodity cell-phones to automatically discover new map semantics to enrich digital maps. Our system depends only on time- and location-stamped inertial sensors measurements, which have a low-energy profile for both road semantics estimation and accurate localization, removing the need for the energy-hungry GPS. For example, a phone going inside a tunnel will experience a drop in the cellular signal strength. This can be leveraged to detect the tunnel location. \system{} uses a classifier-based approach based on the multi-modal phone sensor traces from inside cars to detect tunnels and other road semantics such as bridges, traffic calming devices (e.g. bumps, cat-eyes, etc), railway crossings, stop signs, and traffic lights; In addition, it uses pedestrians' phone sensor traces to detect map semantics like underpasses (pedestrian tunnels), footbridges (pedestrian bridges), crosswalks, stairs, escalators, and number of lanes.

We present the \system{} system architecture as well as the details of its components. Implementation of \system{} over different android phones shows that we can detect different map features accurately with 3\% false positive rate and 6\% false negative for in-vehicle traces, and  2\% false positive rate and 3\% false negative rate for pedestrian traces. In addition, \system{} can detect the location of the detected features accurately to within 2m using as few as 15 samples without using the GPS chip. This comes with a low power consumption of 23mW, which is 50\% less than GPS when run at a 1 minute duty cycle.

In summary, our contributions are three-fold:
\begin{itemize}
    \item We present the \system{} architecture to automatically crowdsense and identify map semantics from available sensor readings without inferring any overhead on the user and with minimal energy consumption. 
    \item We provide a framework for extracting the different map features from both pedestrian and in-vehicle traces.
    \item We implement \system{} on Android devices and evaluate its accuracy and energy-efficiency in a typical city.
\end{itemize}

The rest of the paper is organized as follows: Section \ref{sec:sys_ov}
presents the system overview. We give the details of extracting the road semantic information from phone sensors with pedestrians and in vehicles in sections \ref{sec:ped_sem} and \ref{sec:veh_sem} respectively.  Section~\ref{sec:eval} provides the implementation and
evaluation of \system{}. Section~\ref{sec:rel_work} discusses related work. Finally, Section \ref{sec:conc} provides concluding remarks and give directions for future work.
\begin{figure}[!t]
\centering
\includegraphics[width=0.85\linewidth]{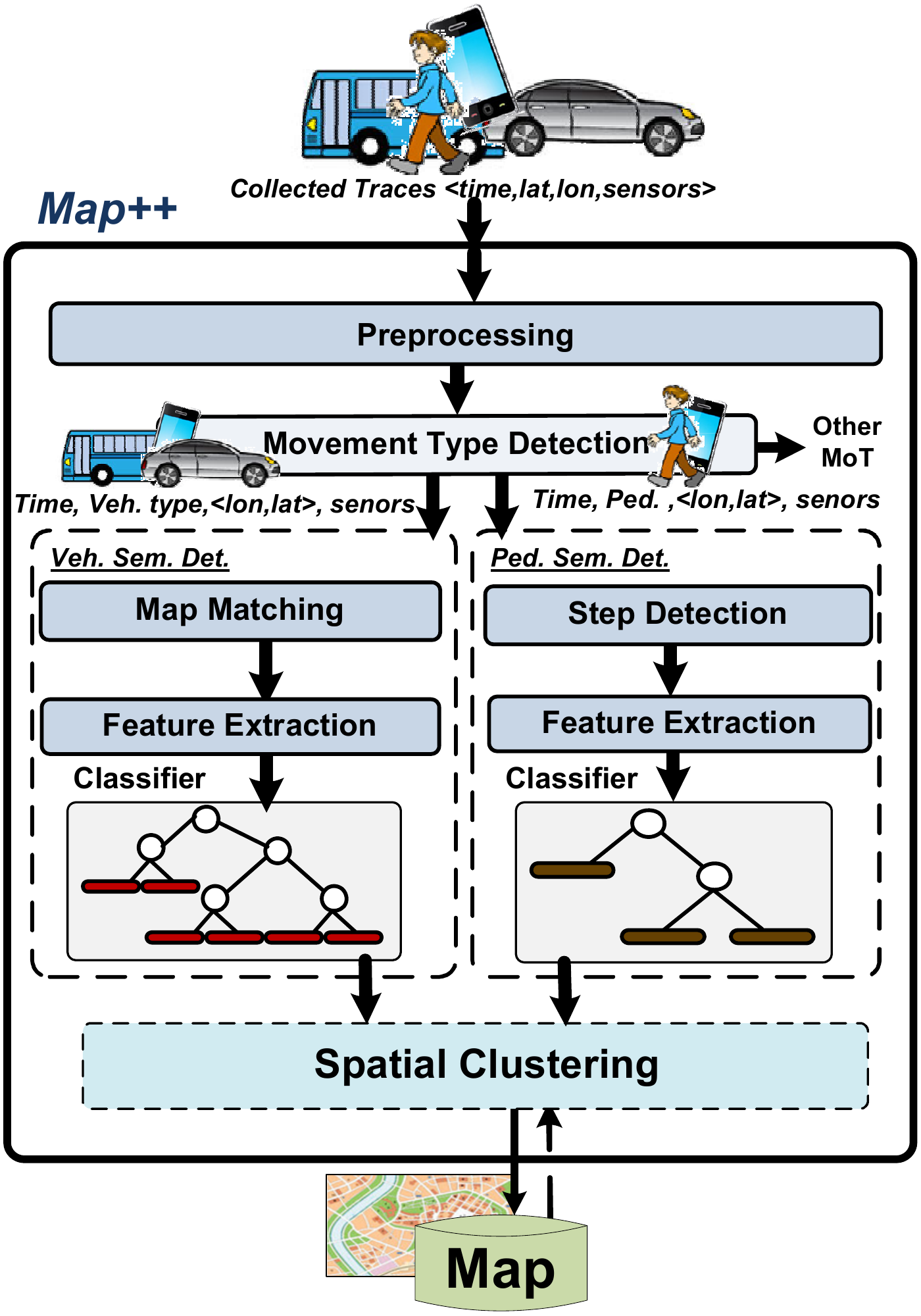}
\caption{The \system{} system architecture.}
\label{fig:sys_arch}
\end{figure}

\section{System Overview}\label{sec:sys_ov}
Figure~\ref{fig:sys_arch} shows the \system{} system architecture. \system{} is based on a crowdsensing approach, where cell phones carried by users submit their data to the \system{} service running in the cloud in a way transparent to the user. The data is first processed by \system{} to reduce the noise. Then, the user mode of transportation is classified to separate pedestrian and in-vehicle traces from other modes of transportation. \system{} has two core components: one for extracting map features from in-vehicle traces and the other for extracting the map features from pedestrian traces. \system{} takes a classifier approach to determine the different road semantics based on extracted features from the collected sensor traces. We give an overview of the system architecture in the following subsections and leave the details for the semantics detection to sections \ref{sec:ped_sem} and \ref{sec:veh_sem}.

\subsection{Traces Collection}
The system collects time-stamped and location-stamped traces along with sensor measurements. These include available inertial sensors (such as accelerometer, gyroscope and magnetometer) as well as cellular network information (associated cell tower ID and its received signal strength (RSS), plus neighboring cell towers and their associated RSS). These sensors have a low cost energy profile and they are already running all the time during the standard phone operation to maintain cellular connectivity or to detect phone orientation changes. Therefore, using them consumes zero extra energy. 

\system{} requires accurate location information while consuming low power; GPS, GSM, and WiFi fingerprinting localization techniques~\cite{GAC,cellsense,ibrahim2011hidden,cellsense2,wifiacc,ibrahim2013enabling} fail to provide both. Hence, \system{} leverages our \emph{Dejavu} energy-efficient and accurate localization system \cite{aly2013dejavu} that can provide accuracy better than GPS in urban conditions with much lower energy consumption. To achieve this, \emph{Dejavu} uses a dead-reckoning approach based on the low-energy phone inertial sensors. However, to reduce the accumulated error in dead-reckoning, \emph{Dejavu} leverages the \textbf{\emph{amble and unique}} physical and logical landmarks in the environments; such as turns, curves, and cellular signal anomalies; as error resetting opportunities.  \emph{Dejavu} can achieve a median distance error of 8.4m in in-city driving conditions and 16.6m in highway driving conditions with a 347.4\% enhancement in energy-consumption compared to the GPS. Therefore, \system{} energy efficiency is based on \emph{Dejavu}'s energy-efficient localization and using the inertial and cellular sensors information for its analysis. We quantify this in Section\ref{sec:eval}.

\subsection{Preprocessing}
This module is responsible for preprocessing the raw sensor measurements to reduce the effect of (a) phone orientation changes and (b) noise and bogus changes, e.g. sudden breaks, or small changes in the direction while moving. To handle the former, we transform the sensor readings from the mobile coordinate system to the world coordinate system leveraging the inertial sensors.
To address the latter, we apply a low-pass filter to the raw sensors data using local weighted regression to smooth the data\cite{cleveland1988locally}. 

\subsection{Transportation Mode Detection}
\system{} is designed to detect two main classes of map semantics; in-vehicle and pedestrian as well as to filter other classes, such as train traces. We start by filtering users inside buildings. Different approaches have been proposed in literature based on the different phone sensors \cite{ravindranath2011improving,krumm2004tempio,zhou2012iodetector}. \system{} uses the IODetector\cite{zhou2012iodetector} approach due to its accuracy and low-energy profile.

Similarly, transportation mode detection for outdoor users has been thoroughly studied in the literature \cite{trnsp_mode_1,trnsp_mode_2,trnsp_mode_3}.
We follow the approach proposed by \cite{trnsp_mode_2} that provides high accuracy of differentiation between the different transportation modes based on the energy-efficient inertial sensors. The technique starts by segmenting the location traces\cite{zheng2008learning} using velocity and acceleration upper bounds. Then the following features are used to classify each segment: The stopping rate, the heading and velocity change rate, the segment length, the $i^{th}$ maximum velocity and acceleration, average velocity, and velocity variance. A decision tree classifier is applied to identify the transportation mode for each segment.

Once the mode of transportation is detected, a map-matcher \cite{tang2012efficient} is applied to the in-vehicle traces to map the estimated locations to the road network to reduce the localization error. Similarly, the UPTIME step detection algorithm\cite{alzantot2012uptime} that takes into account the different users' profiles and  gaits is applied to the pedestrian acceleration signal to detect the user steps. In both cases, features are extracted from the traces to prepare for the road semantic classification step.
\subsection{Map Semantics Extraction}
There are a large number of road semantic features that can be identified based on their unique signature on the different phone sensors.
\system{} uses a tree-based classifier to identify the different semantics as detailed in Section~\ref{sec:ped_sem} (pedestrian-based semantics)  and Section~\ref{sec:veh_sem} (vehicle-based semantics).

\subsection{Road Semantic Features Location Estimation}
Whenever a road semantic feature is detected by the semantic detection modules (in-vehicle or pedestrians), \system{} needs to determine whether it is a new instance of the road feature or not as well as determine its location. 

To do this, \system{} applies spatial clustering for each type of the extracted semantics. It uses density-based clustering algorithms (DBSCAN\cite{ester1996density}). DBSCAN has several advantages as the number of clusters is not required before carrying out clustering; the detected clusters can be represented in an arbitrary shape; and outliers can be detected. The resulting clusters represent map features. The location of the newly discovered semantics is the weighted mean of the points inside their clusters. We weight the different locations based on their accuracy reported by \emph{Dejavu}: In \emph{Dejavu}, the longer the user trace from the last resetting point, the higher the error in the trace\cite{aly2013dejavu}. Therefore, shorter traces have better accuracy.  When a new semantic is discovered, if there is a discovered map feature within its neighborhood, we add it to the cluster and update its location. Otherwise, a new cluster is created to represent the new road feature. To reduce outliers, a semantic is not physically added to the map until the cluster size reaches a certain threshold.

\subsection{Practical Considerations}
Sensor specifications are different from one phone manufacturer to another, which leads to different sensor readings for the same map feature. To address this issue, \system{} applies a number of techniques including use of scale-independent features (e.g. peak of acceleration) and combining a number of features for detecting the same semantic feature.

\system{} does not also require real-time sensor data collection; it can store the different sensor measurements and opportunistically upload them to the cloud for processing; allowing it to save both communication energy and cost. This is outside the scope of this paper.

\section{Pedestrian Traces Semantic Detection Module}\label{sec:ped_sem}
\begin{figure*}[!t]
\centering
\includegraphics[width=0.77\linewidth]{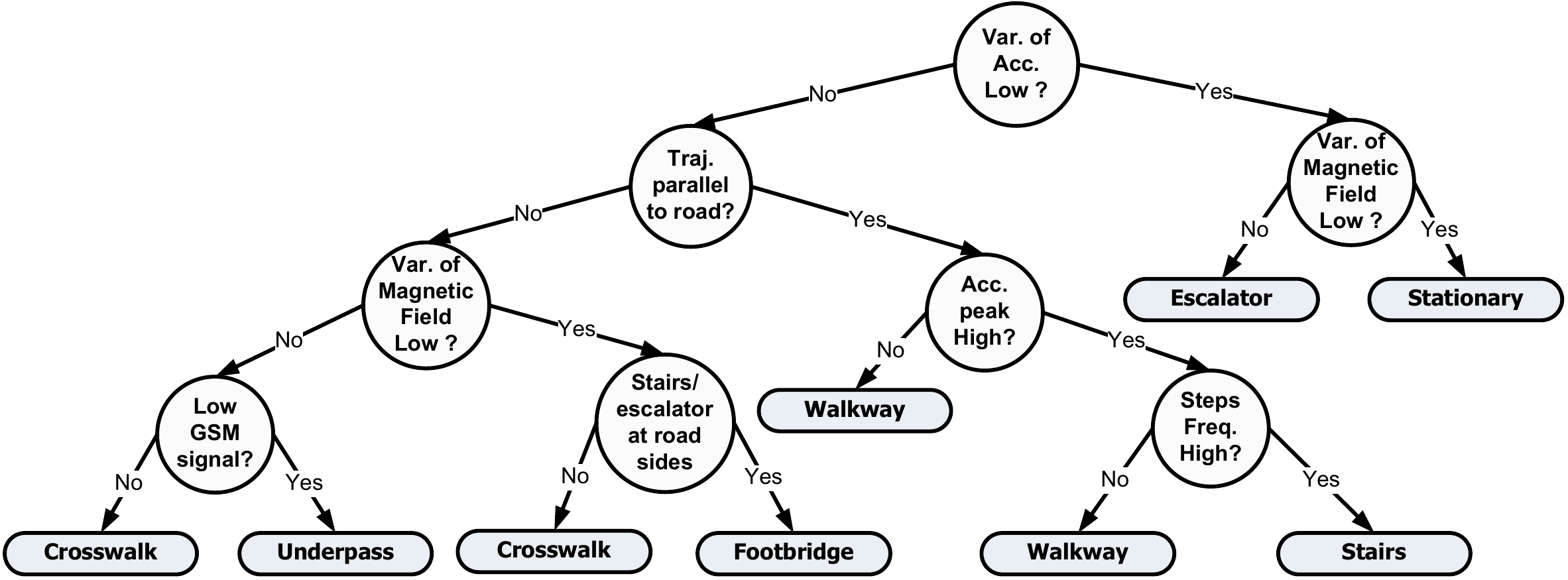}
\caption{A decision tree classifier for detecting different map features from {\bfseries pedestrian} traces.}
\label{fig:ped_sem_state}
\end{figure*}

To determine the different road semantics, \system{} applies a decision tree classifier to the extracted features from the pedestrian traces. Figure \ref{fig:ped_sem_state} shows the decision tree classifier used to extract the different semantic map features from the pedestrian traces. We give the details of the classifier features that can differentiate the different semantic road features (underpasses, stairs, escalators, footbridges, crosswalks, and number of lanes) in the next subsections.

\subsection{Underpasses (Pedestrian Tunnels)}
\begin{figure}[!t]
\centering
    \subfigure[Magnetic field sensor]{
      \includegraphics[width=\figscale\linewidth]{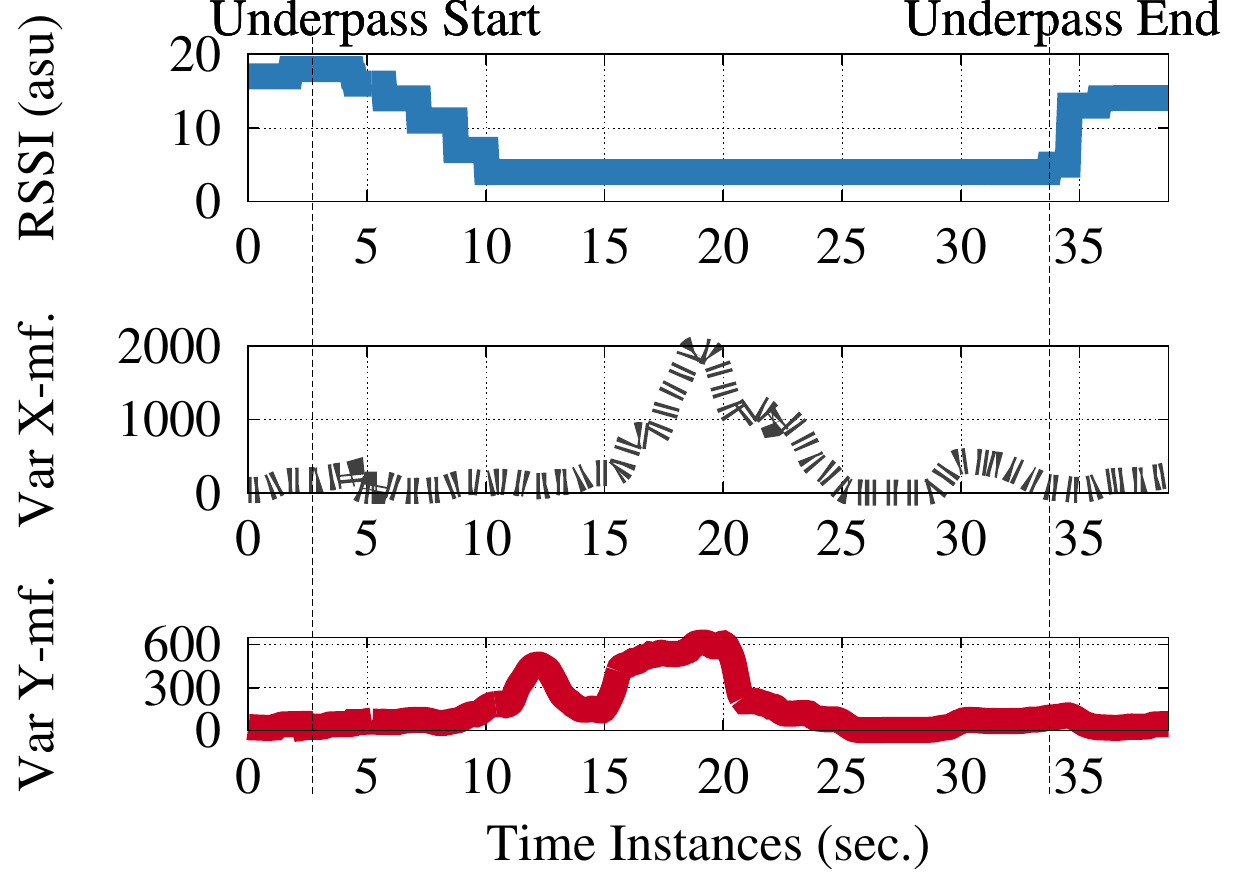}
      \label{fig:underpass}
    }
    \subfigure[Map]{
      \includegraphics[width=\figscale\linewidth,height=0.1\textheight]{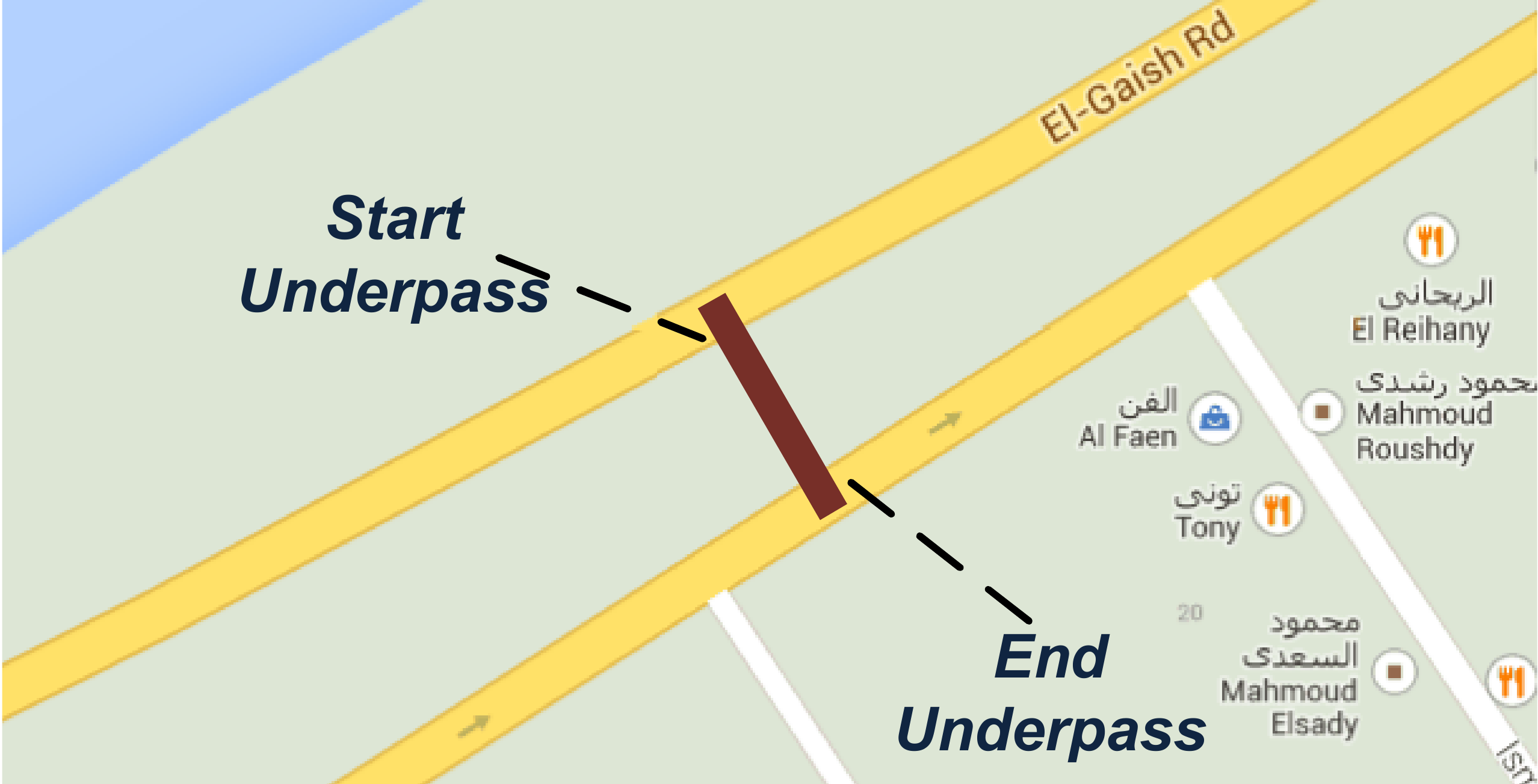}
  \label{fig:underpass_map}
    }
\caption{The effect of walking in an underpass on the cellular signal strength and the ambient magnetic field. Typically, the cell-phone will experience a drop in the cellular signal and also a high variance in the magnetic field in both sides of the tunnels.}  \label{fig:tunnel_ped}
\end{figure}

Underpasses or pedestrian tunnels are specially constructed for pedestrians beneath a road or railway, allowing them to reach the other side. A pedestrian trace crossing a road may be a crosswalk (e.g. zebra lines), a bridge, or an underpass. We identify the underpasses from other classes by their unique features: Walking inside an underpass, a cell-phone will experience a drop in the cellular signal and also a high variance in the magnetic field around it (both Y and X axes) due to metals and electricity lines inside the tunnel (Figure~\ref{fig:tunnel_ped}).

\subsection{Stairs}

\begin{figure}[!t]
\centering
\includegraphics[width=\figscale\linewidth]{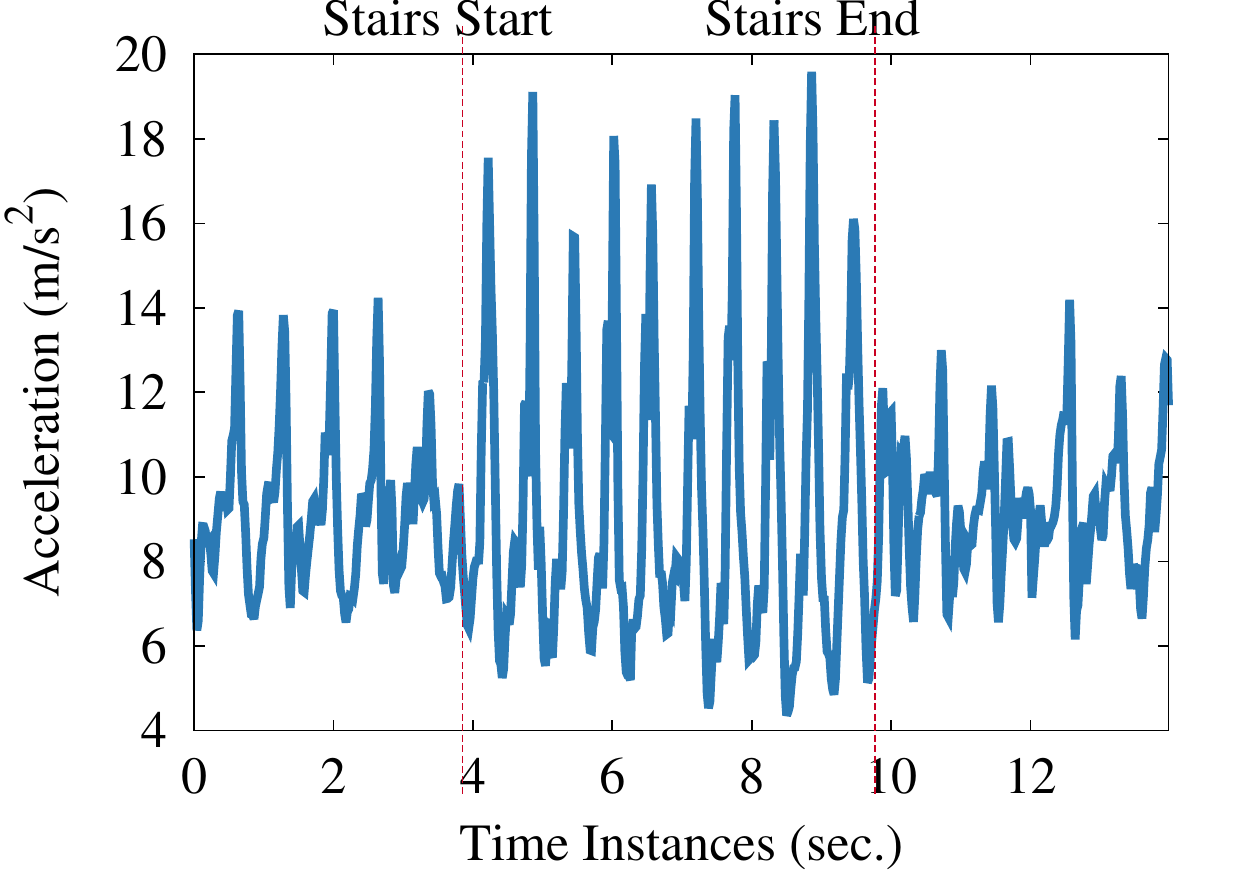}
\caption{Step pattern while walking versus the step pattern when going down stairs. The stairs are clear from the walking pattern as they have a higher peak in comparison to the normal walking peaks. The number of steps is detected with a simple peak detector.}
\label{fig:stairs}
\end{figure}

\begin{figure}[!t]
\centering
\includegraphics[width=0.75\linewidth]{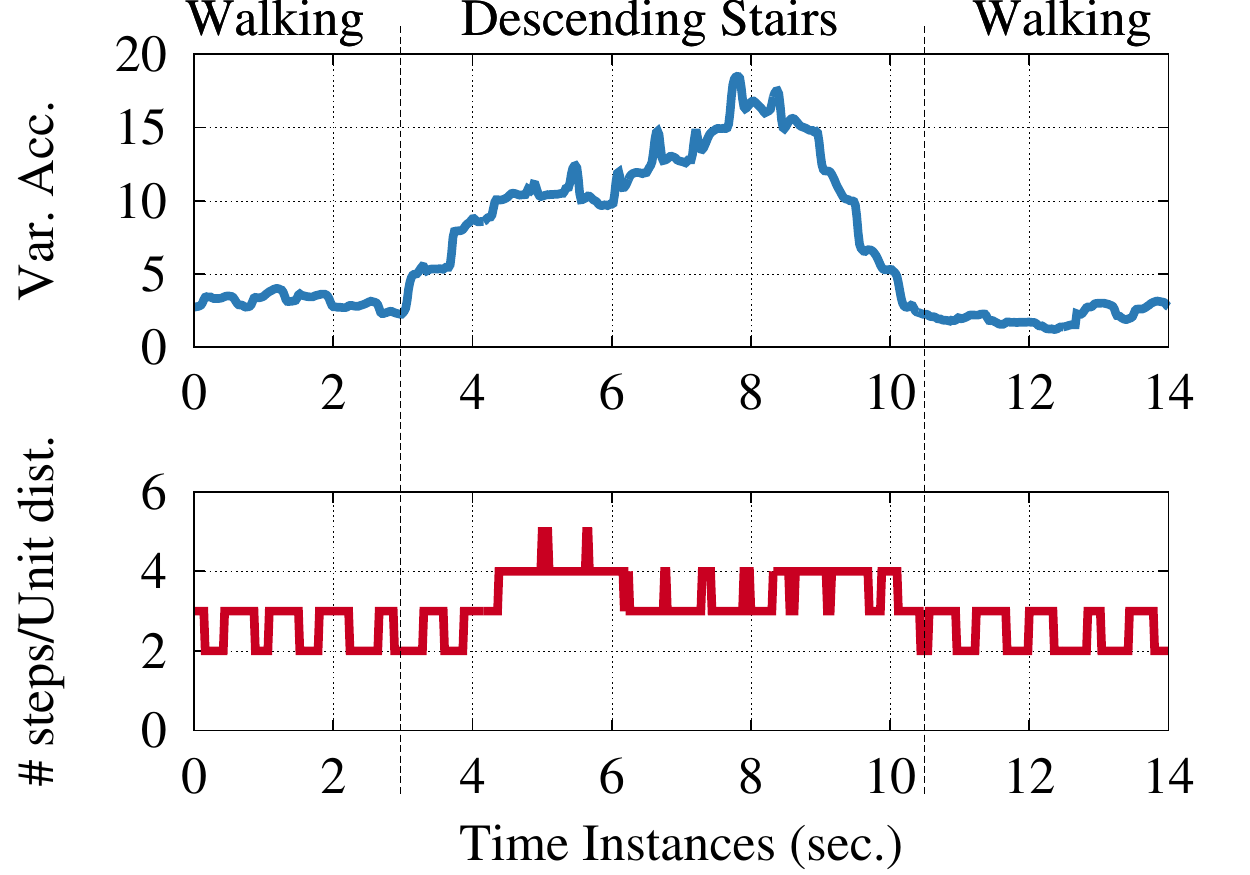}
\caption{The figure compares the effect of descending stairs to the effect of walking on the acceleration variance and the steps frequency. Descending stairs leads to a higher magnitude in acceleration and higher frequency.}
\label{fig:stairsignature}
\end{figure}

Furthermore, when ascending or descending stairs, the frequency of steps, detected by a simple peak detector (Figure~\ref{fig:stairs}), within the unit distance increases since the user is moving vertically. When descending stairs, the gravity force affecting the person will lead to a higher peak in acceleration, and hence higher variance, as compared to walking (Figure~\ref{fig:stairsignature}). The number of steps can be used, e.g., to determine the height of the pedestrian bridge, which is useful for determining the height limits for the vehicles on the road. 

\subsection{Escalators}
\begin{figure}[!t]
\centering
\includegraphics[width=\figscale\linewidth]{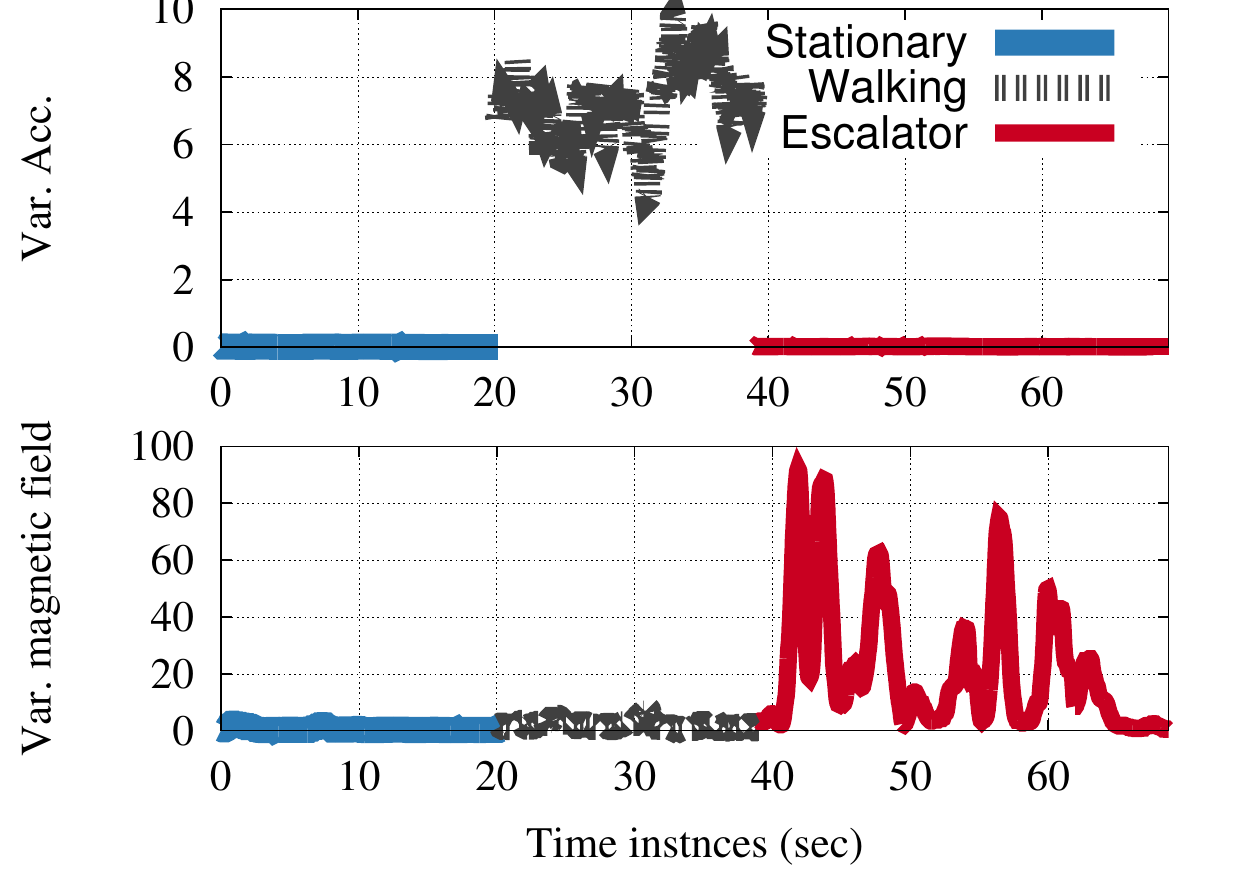}
\caption{Variance of acceleration and ambient magnetic field while being stationary, walking, and using the escalator.}
\label{fig:escalator}
\end{figure}

When using escalators, users typically keep standing while carried by the moving staircase. Therefore, the acceleration variance remains small  compared to walking. However, escalators are often powered by constant-speed alternating current motors, which results in high variance in the magnetic field (Figure~\ref{fig:escalator}).

\subsection{Footbridges (Pedestrian bridges)}
\begin{figure}[!t]
\centering
\includegraphics[width=\figscale\linewidth]{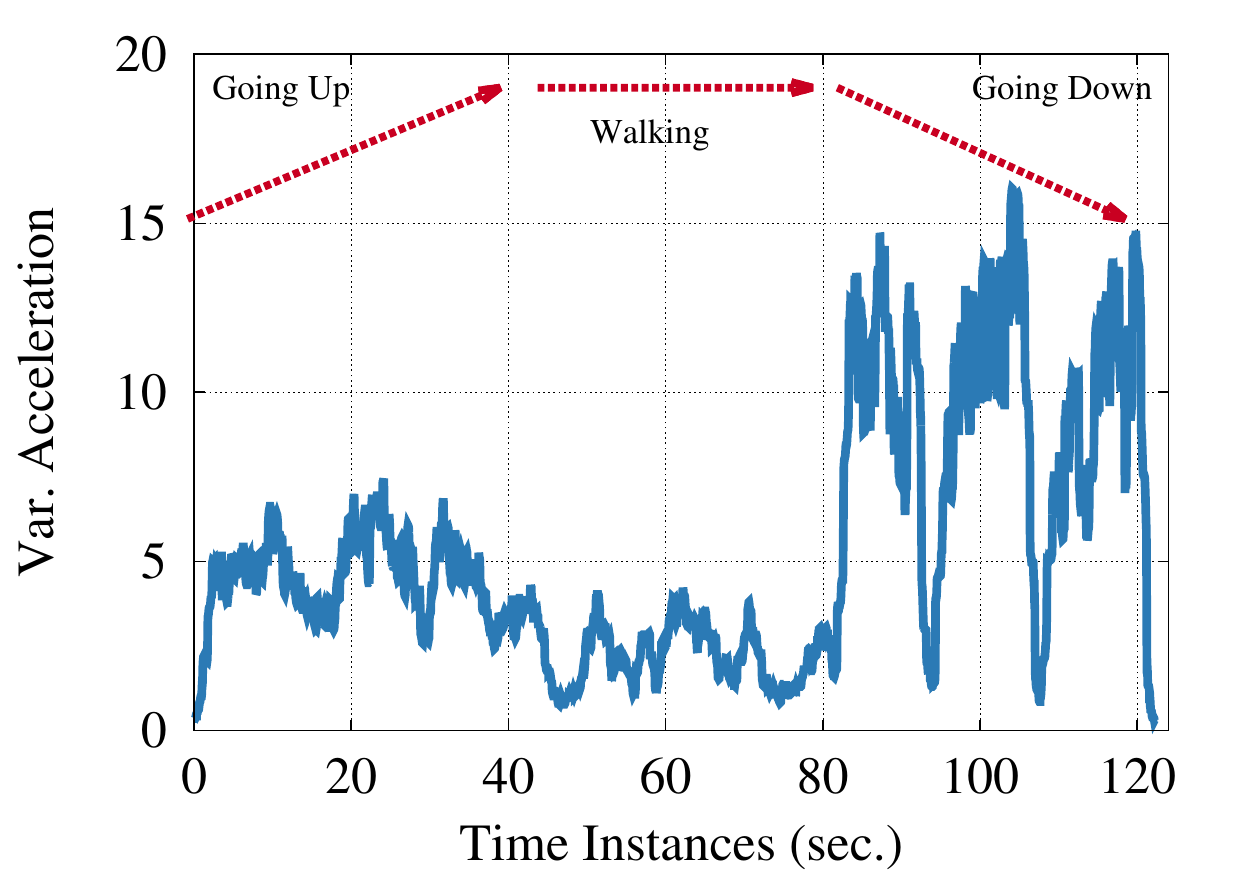}
\caption{Acceleration variance when going up then walking on the footbridge then going down on stairs.}
\label{fig:ped_bridge}
\end{figure}

Similar to underpasses, footbridges allows pedestrians to safely cross roads, railways and rivers. A user crossing a footbridge will use stairs/escalators to ascend and descend. In between, the user will walk the length of the footbridge (Figure~\ref{fig:ped_bridge}).
We separate footbridges from crossroads by detecting the stairs/escalator pattern before/after using them; we separate them from underpasses using the cellular signal  which drops in the underpasses case but not in the footbridge case.

\subsection{Crosswalk}
A crosswalk or pedestrian crossing is a designated point on a road to assist pedestrians wishing to cross, they could be found at intersections and busy roads. Crosswalks could be detected from pedestrian traces; a point on the road where users cross to another road without footbridges or underpasses.

\subsection{Number of Lanes}
Road capacity is an important feature for applications like traffic estimation  and evacuation route planning. We detect the number of lanes from the road width divided by the average lane width in the region. Road lanes typically have an accepted width range set by an authority for every country\cite{ustrsnp}.
We detect road width from the length of the crosswalk, the footbridge, or the underpass that helps pedestrian cross it. Since users can have different patterns for crossing the road, we found that using the minimum road length leads to an accurate estimate for the number of lanes.

\section{In-Vehicle Traces Semantic Detection Module}\label{sec:veh_sem}

\begin{figure}[!t]
\centering
\includegraphics[width=\linewidth]{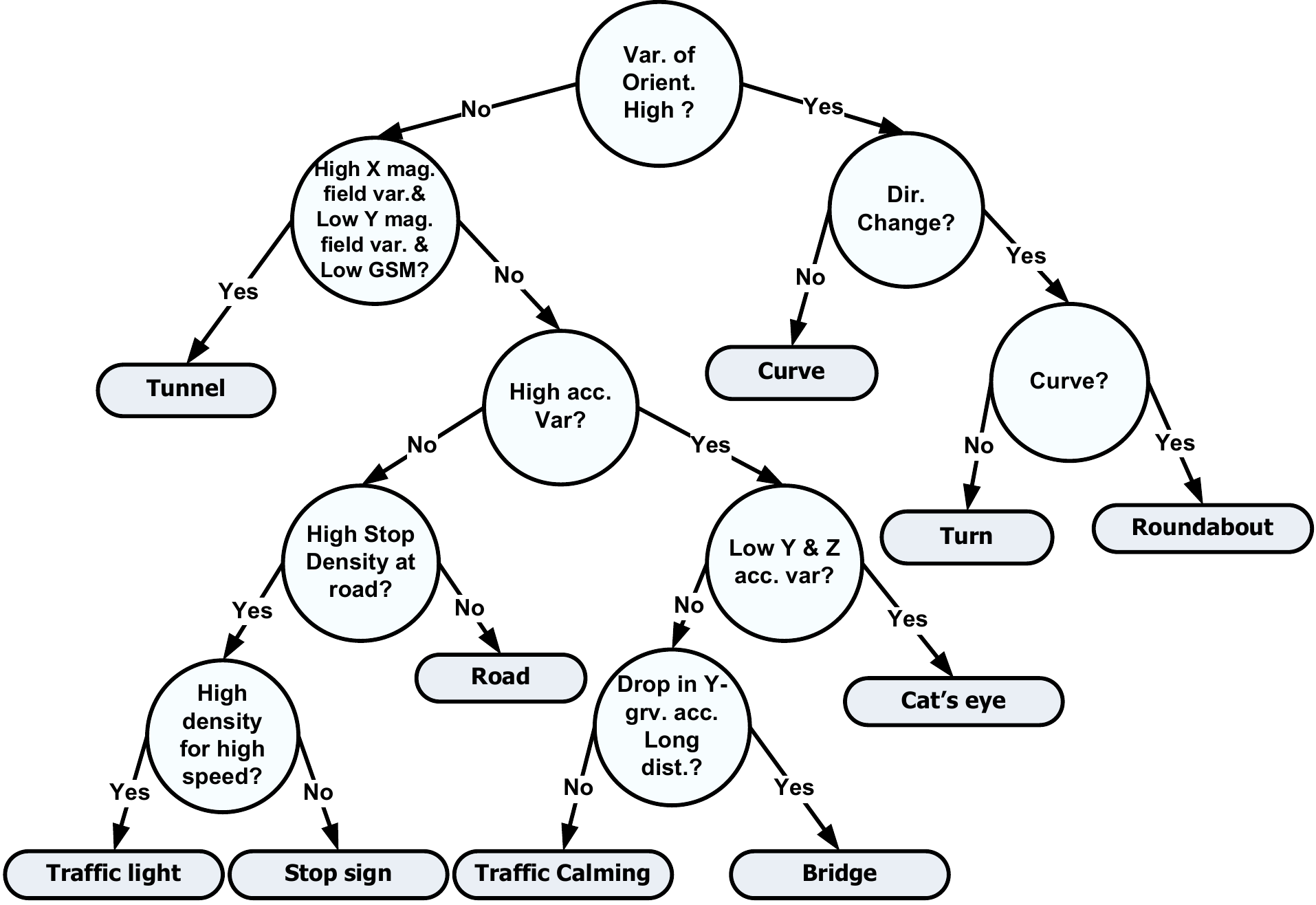}
\caption{A decision tree classifier for detecting different map features from {\bfseries in-vehicle} traces.}
\label{fig:veh_sem_state}
\end{figure}

We extract the different map semantic features from the traces collected by the in-vehicle users.  Figure \ref{fig:veh_sem_state}, shows the tree classifier used to detect the different semantics.

\subsection{Tunnels}
Similar to the underpass case, a car going inside a tunnel will typically experience an attenuated cellular signal. We also noticed a large variance in the ambient magnetic field in the x-direction (perpendicular to the car direction of motion) while the car is inside the tunnel. This is different from the underpass case, where there is no smooth ramp at the end and hence both the x and y magnetic fields are affected. Therefore, car tunnels have a low variance in the y-axis (direction of car motion) magnetic field.

\subsection{Bridges}
\begin{figure}[!t]
\centering
\includegraphics[width=0.75\linewidth]{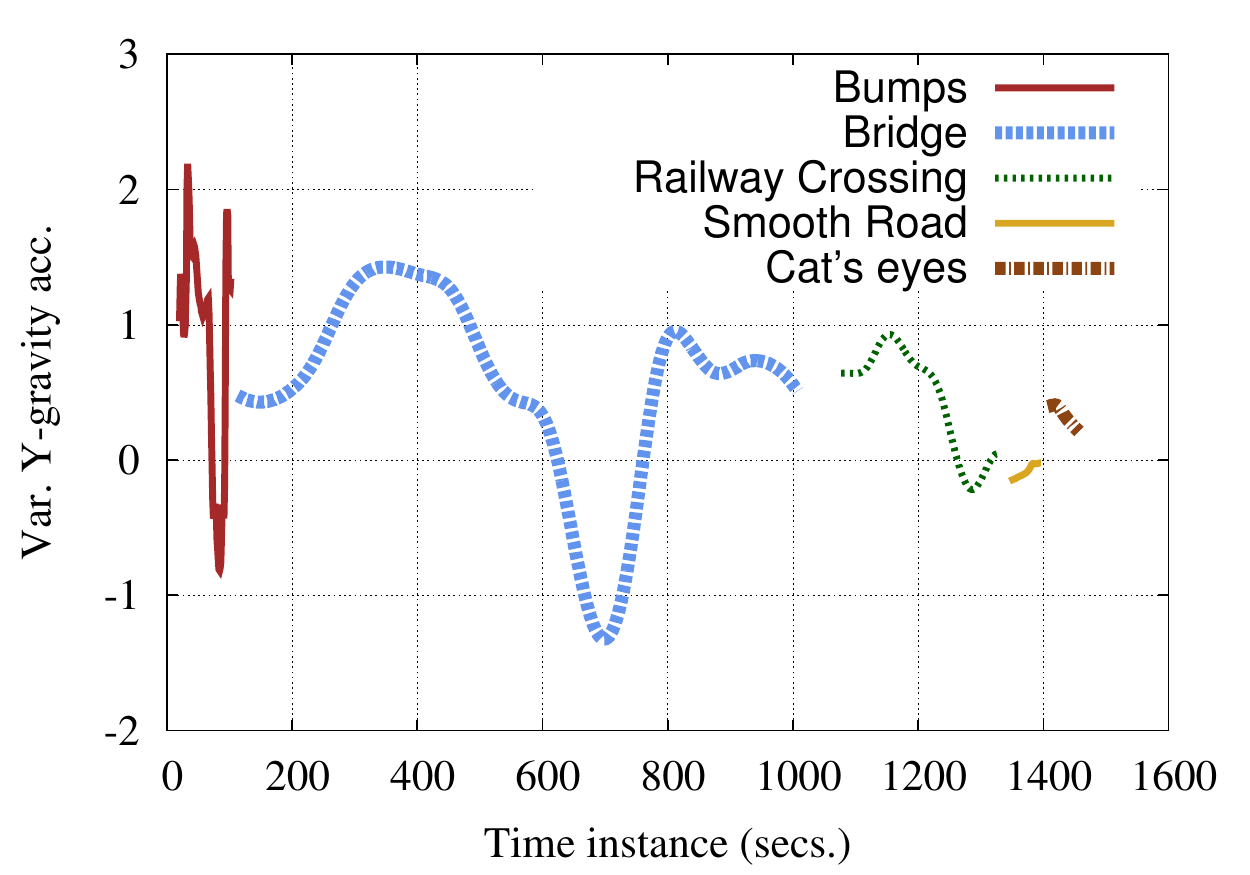}
\caption{Effect of the different map features on the Y-axis gravity acceleration.}
\label{fig:bridge}
\end{figure}

Bridges cause the car to go up at the start of the bridge and then go down at the end of the bridge. This is reflected on the Y-gravity or Z-gravity acceleration (Figure~\ref{fig:bridge}). Although other classes, such as bumps, cause the same effect (Y or Z gravity acceleration going up then down), bridges are unique in having this effect over a longer distance~\cite{aly2013dejavu}. The bridge is detected at its end. Note that after detecting the end of the bridge, we could identify its starting point.

\subsection{Traffic Calming Features}
\begin{figure}[!t]
\centering
    \subfigure[X variance.]{
      \includegraphics[width=0.76\linewidth]{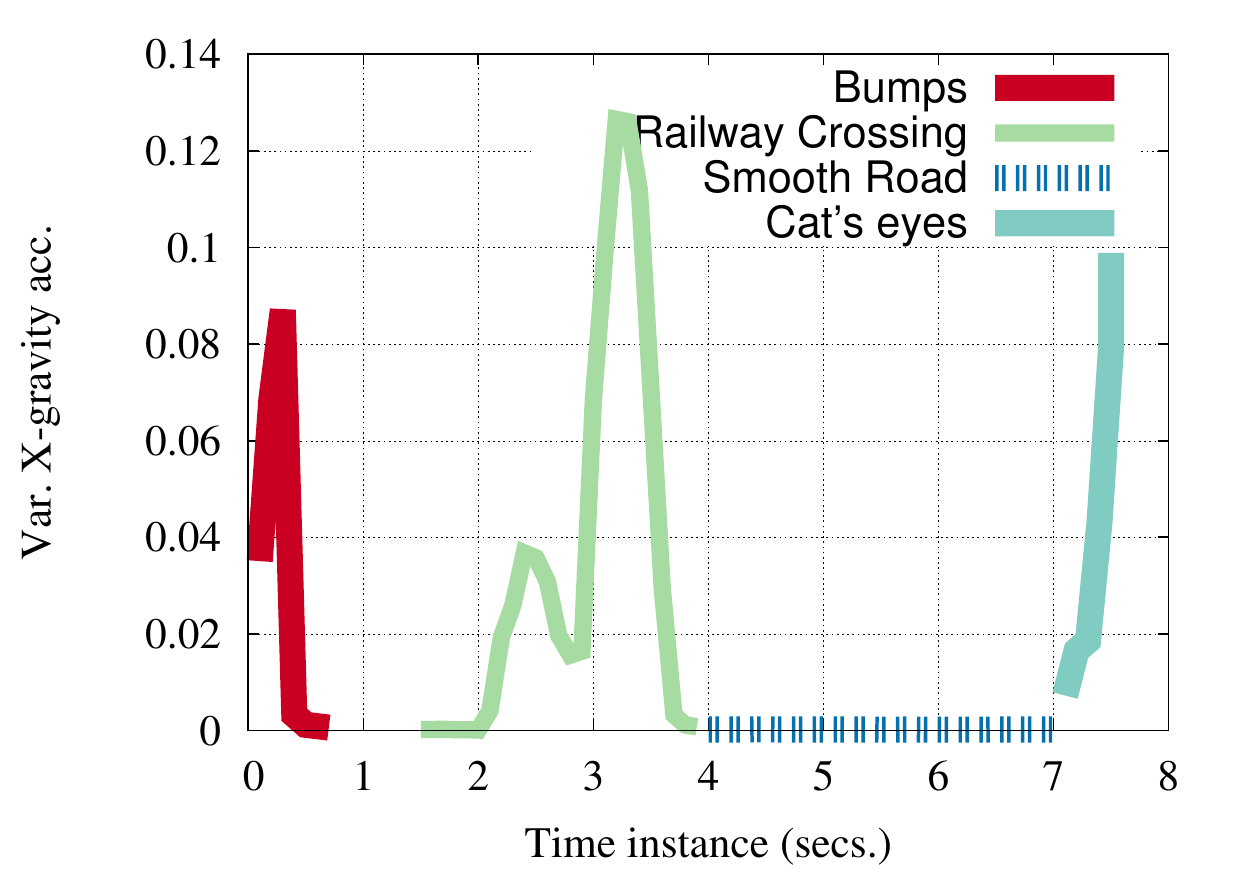}
      \label{fig:road_x}
    }
    \subfigure[Y variance]{
      \includegraphics[width=0.76\linewidth]{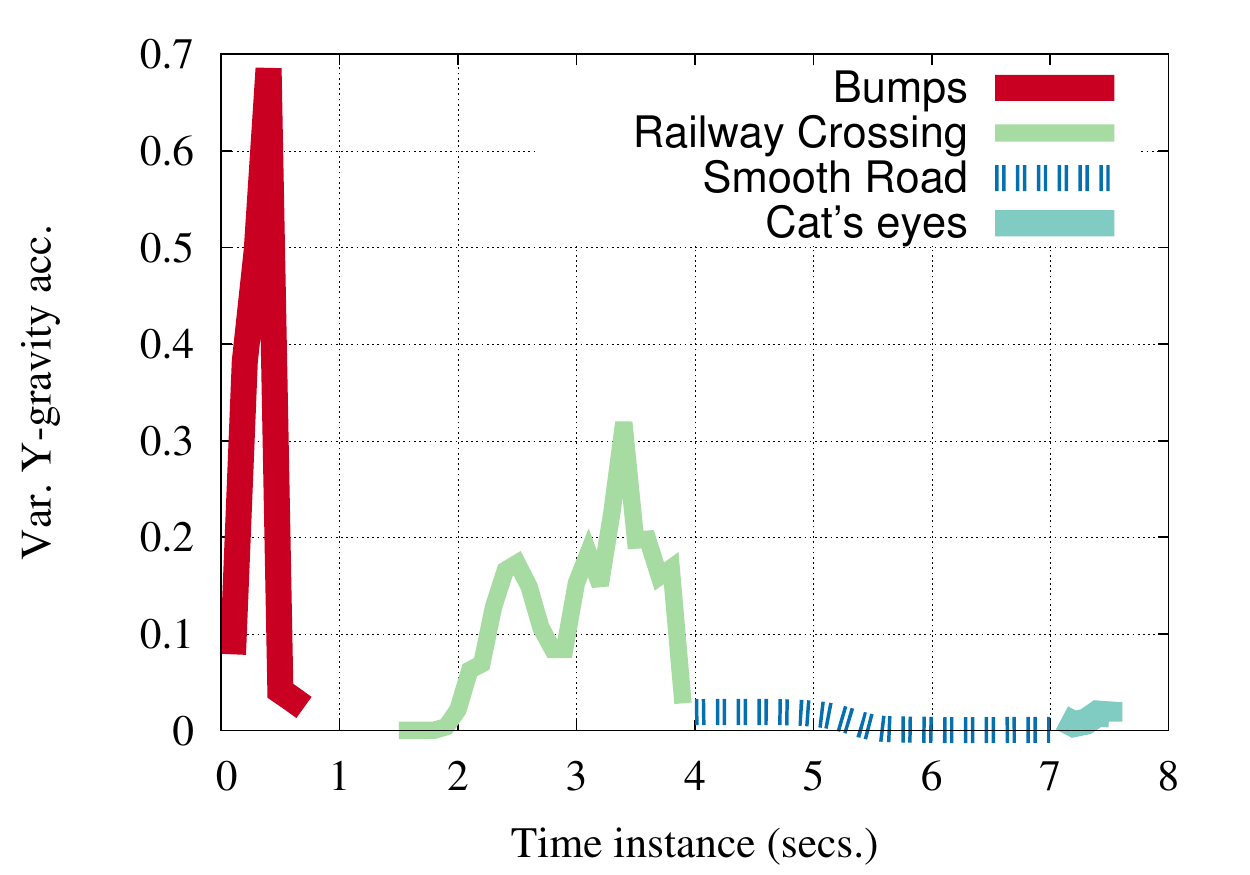}
  \label{fig:road_y}
    }
    \subfigure[Z variance]{
      \includegraphics[width=0.76\linewidth]{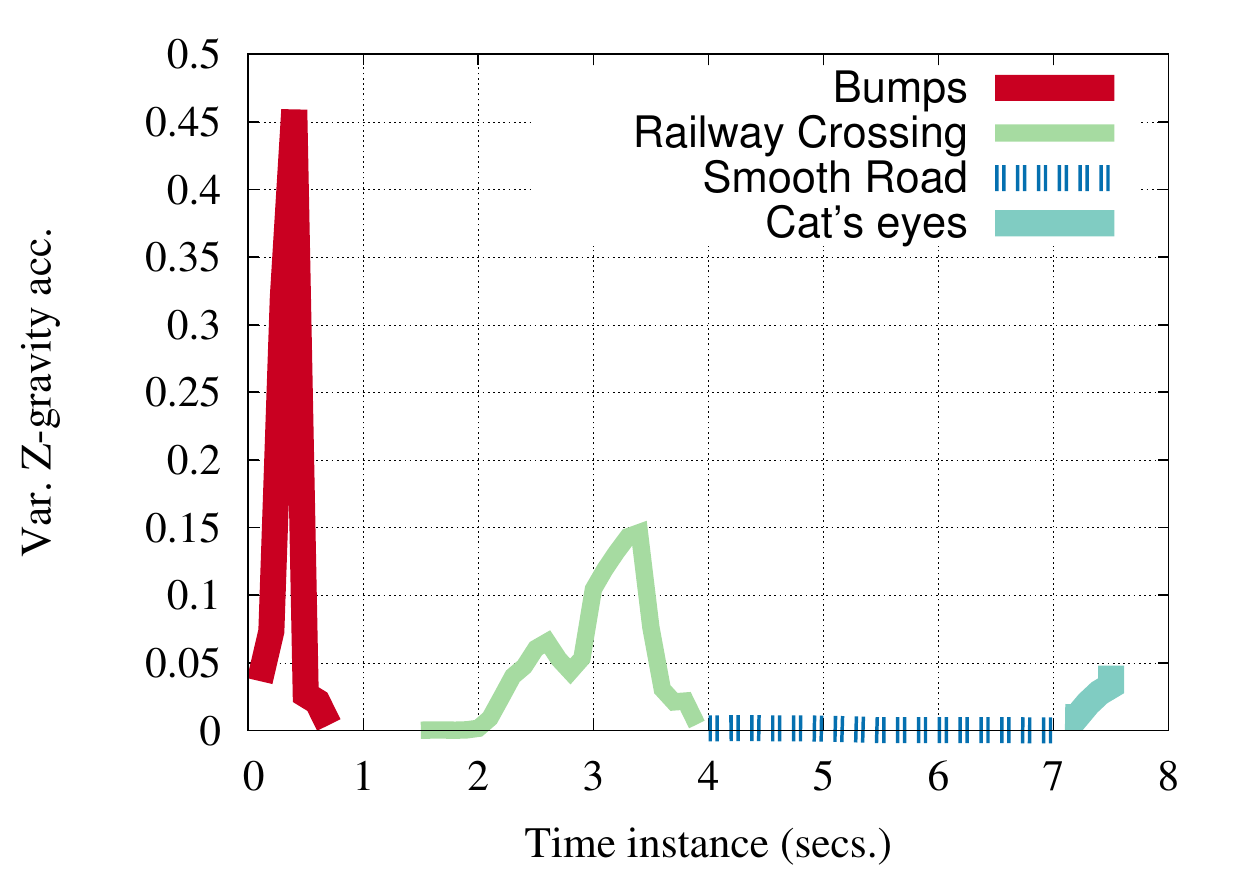}
  \label{fig:road_z}
    }
  \caption{Effect of different traffic calming devices on the X, Y, and Z gravity acceleration variance in comparison with smooth road and railway crossing. Cat's eyes have the lowest Y and Z variance, bumps have the highest Y and Z variances, while the railway crossing has a medium Y and Z variance. }
  \label{fig:road}
\end{figure}

Different traffic calming techniques like bumps, speed humps, and cat's eyes all cause the car to move up then down similar to bridges, affecting all gravity acceleration axes. However, unlike bridges, all these classes affect the gravity acceleration over a small distance. To further separate these classes, we employ other sensors using the approach in~\cite{aly2013dejavu} as follows (Figure~\ref{fig:road}):

{\bfseries Vertical deflection devices (e.g., speed bumps, humps, cushions, and speed tables):} As the vehicle hits such devices, large spikes in variance in the Y-axis and Z-axis gravity acceleration are sensed compared to the other classes.

{\bfseries Cat's eyes:} Unlike other road anomalies, the cat's eyes structure does not cause the car moving above them to have high variance in the Y or Z-axis gravity acceleration.

\subsection{Railway Crossing}
Railway crossings leads to a medium variance in the Y-axis and Z-axis gravity acceleration over a longer distance compared to other road anomalies~\cite{aly2013dejavu}. In addition, they cross a railway if available on the map.

\subsection{Roundabouts and Intersections}
\begin{figure}[!t]
\centering
\subfigure[4-way intersection]{
      \includegraphics[width=0.36\linewidth]{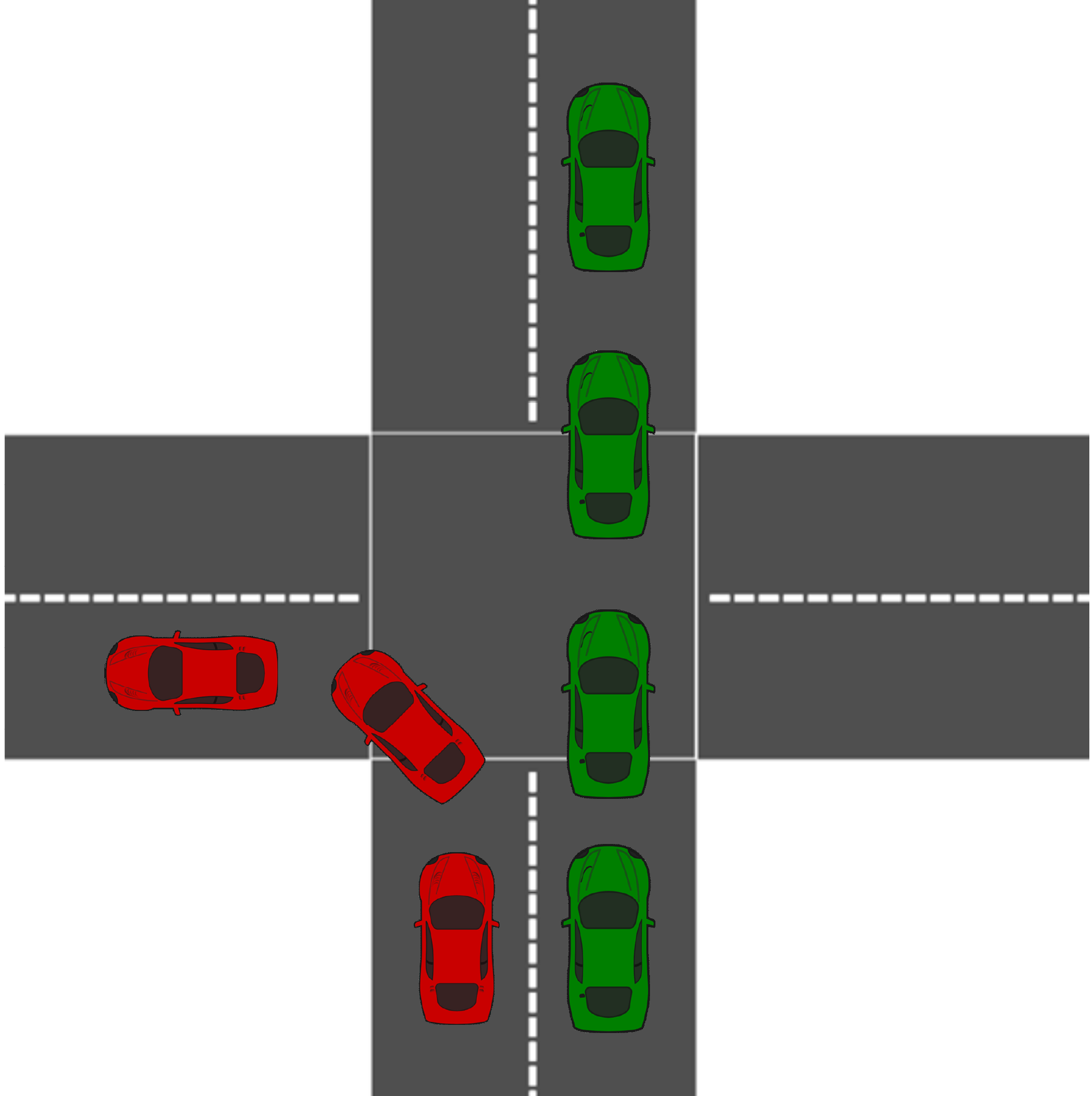}
      \label{fig:inter}
    }
    \subfigure[Roundabout]{
      \includegraphics[width=0.36\linewidth]{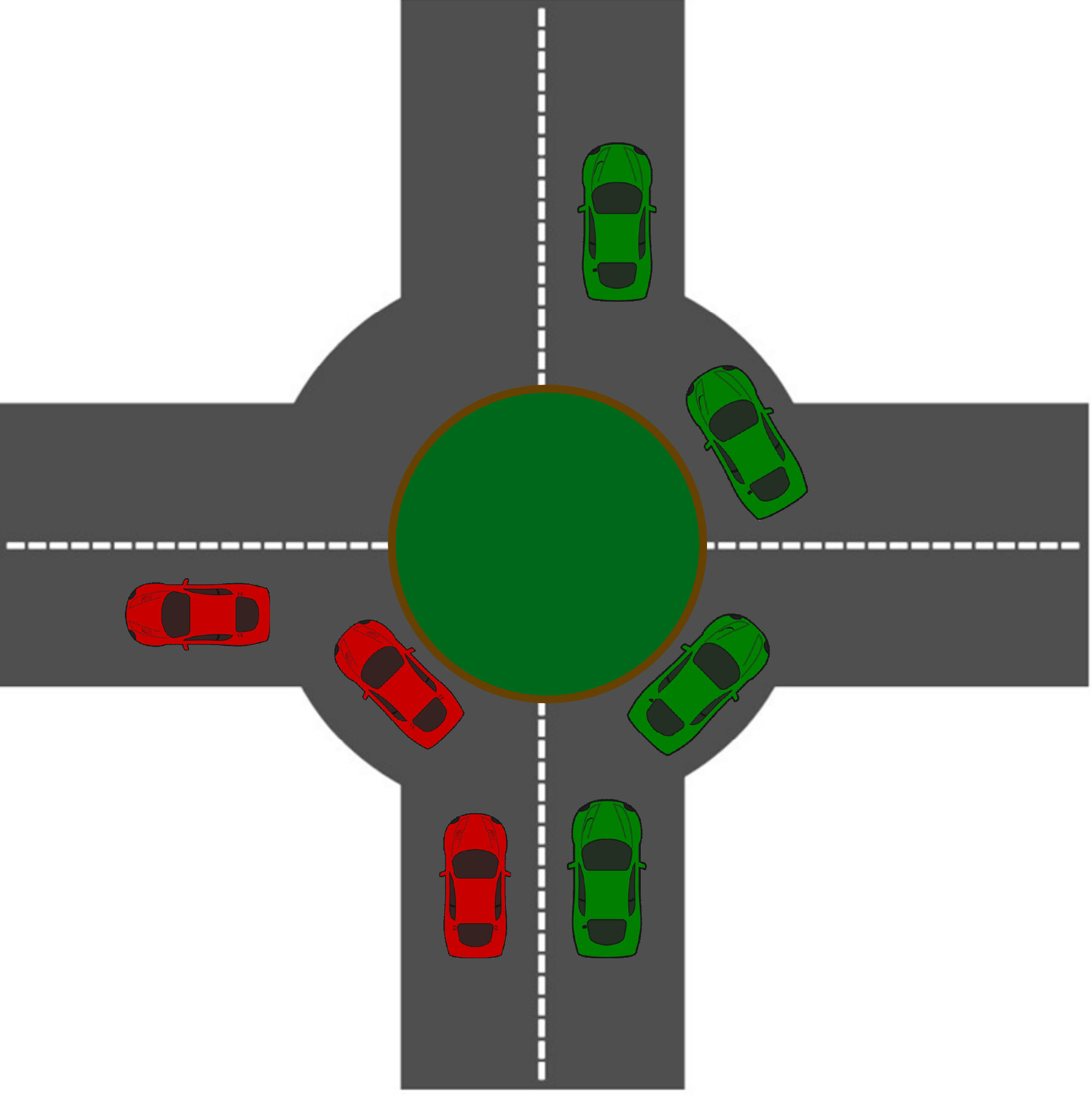}
  \label{fig:round}
    }
\caption{Difference between a roundabout and an intersection in terms of car behavior. }
\label{fig:inter_round}
\end{figure}

A roundabout is a type of circular junction in which road traffic must travel in one direction around a central island. While a four-way intersection are typically two perpendicular crossing roads (Figure~\ref{fig:inter_round}). Roundabouts can be identified as normal crossings by some commercial map services as shown in Figure~\ref{fig:map_comp}.

\begin{figure}[!t]
\centering
    \subfigure[Curve]{
      \includegraphics[width=0.46\linewidth]{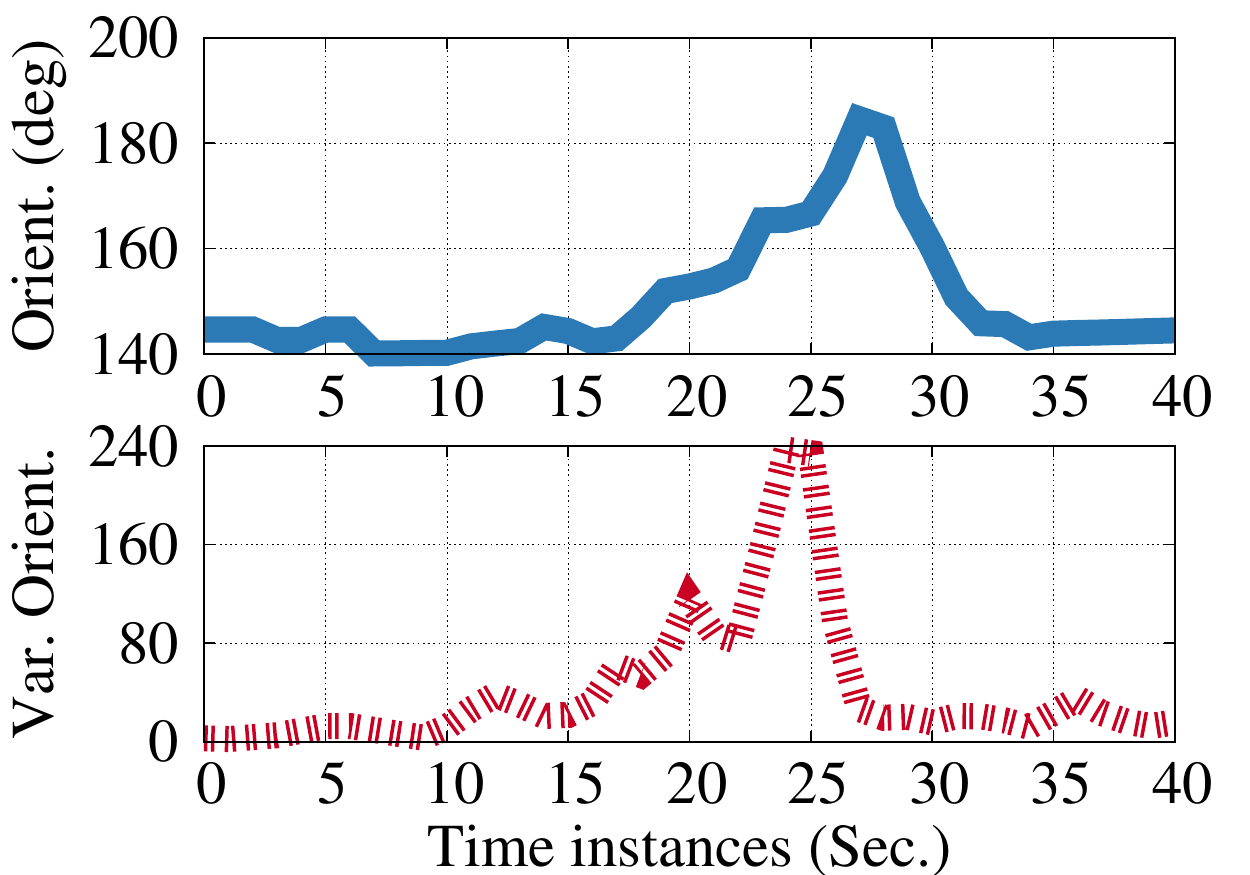}
      \label{fig:veh_curve}
    }
    \subfigure[Turn]{
      \includegraphics[width=0.46\linewidth]{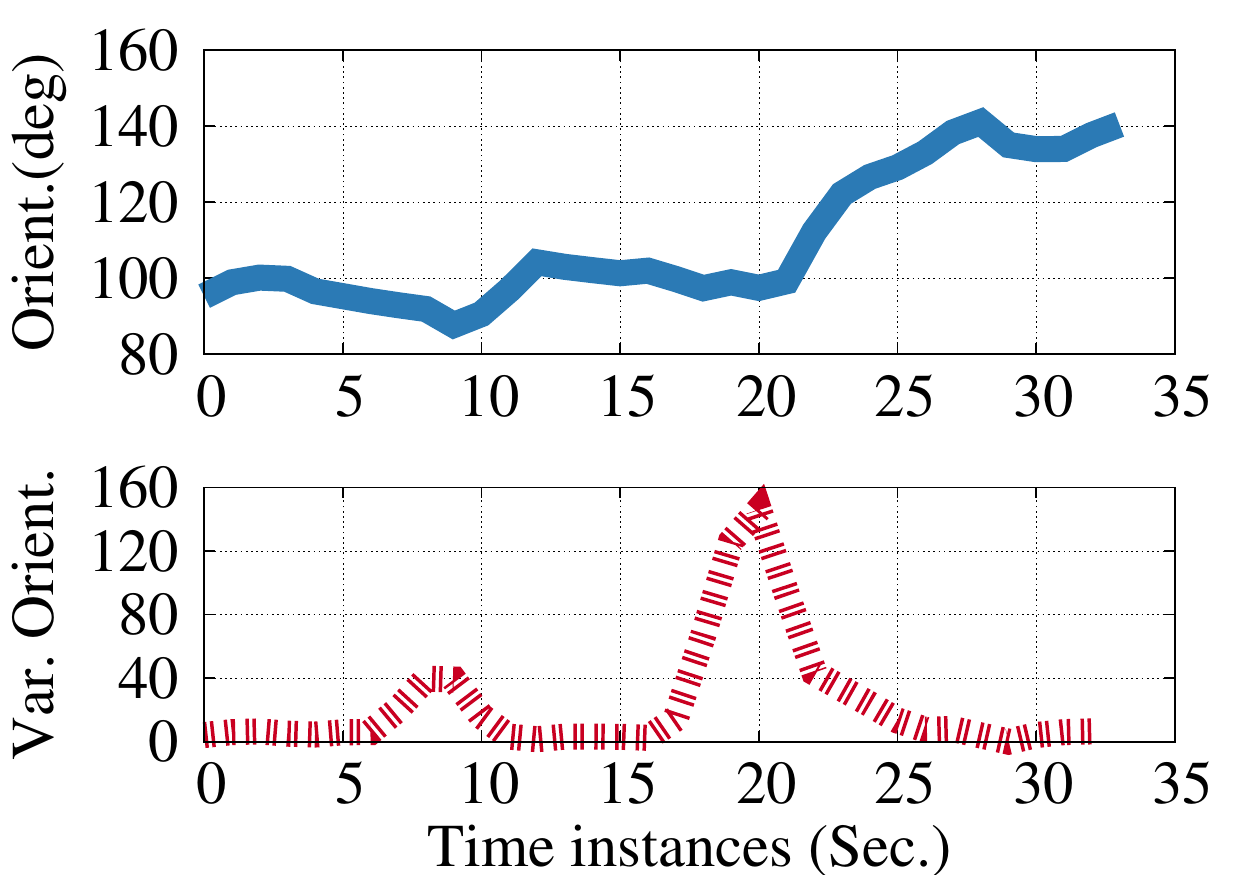}
  \label{fig:veh_turn}
    }
\caption{Moving along the curve the car direction (orientation) changes till the end of the curve, where the car returns to its original direction. This is unlike taking a turn, where direction changes after the end of the curving part remains.}
\label{fig:curve}
\end{figure}

Noting that a four-way intersection will only have sharp 90$^\circ$ turns; while a roundabout will have both turns and curves (Figure \ref{fig:inter_round}), we can leverage the orientation angle sensor to identify the roundabouts by the differences between their start and end orientation angles (Figure~\ref{fig:curve}).

\subsection{Stop Sign and Traffic Light}
Traffic lights and stop signs are often used at road intersections, pedestrian crossings and other locations to control competing flows of traffic. Different approaches were proposed to identify both traffic regulators. We use the approach proposed in \cite{carisi2011enhancing} to detect the location and timing of both stop-signs and traffic lights based on the location traces. From both traffic lights and stop signs characteristics, all traffic passing by a stop sign should stop or at least slowdown (due to driver behavior), while the behavior of vehicles approaching a traffic light depends on if the light is actually red or green. So typically only a fraction of vehicles stops.

They identify a potential stop sign if at least 80\% of the traces slow down at the intersection and identify a potential traffic light if at least 15\% of the traces slow down at the intersection. If all or all but one directions belonging to a given intersection are marked as potential stops, the intersection is identified as stop-sign regulated, and all the ways marked with a potential stops become actual stop signs; and an intersection is considered regulated by a traffic light if half plus one of the incoming directions are marked as potential traffic light.

\section{Evaluation}\label{sec:eval}
We implemented our system on different Android phones. including: HTC Nexus One, Samsung
Galaxy Nexus, Galaxy Tab 10.1, and Samsung Galaxy S Plus. We carried out our experiments on different roads in the city of Alexandria, Egypt. The total distance for in-vehicle traces is 40km (covering an area of about 10km$^2$) and  15km (covering an area of about 1.5km$^2$) for the pedestrian traces. Traces were collected by three subjects. The ground truth for the map features were marked manually. For the location, we used an external bluetooth satellite navigation system that uses both the GPS and GLONASS systems as ground truth and ignored the low-accuracy GPS embedded in the phone.

For the rest of this section, we start by evaluating the semantics detection accuracy for the different roads, the location accuracy for the discovered map features, and the power consumption of \system{} as compared to GPS-based approaches that focus on updating the road segments only \cite{crowdAtlas,cao2009gps,baier2011mapcorrect}.

\subsection{Road Features Detection Accuracy}
Tables \ref{tab:drive_conf} and \ref{tab:ped_conf} show the confusion matrices for detecting the different map semantics from in-vehicle and pedestrian traces, respectively. The tables show that different map features could be detected with small false positive and negative rates due to their unique signatures; we can detect the map semantics accurately with 3\% false positive rate and 6\% false negative rate from in-vehicle traces, and 2\% false positive rate and 3\% false negative rate from pedestrian traces.

\begin{table*}[!t]
\caption{Confusion matrix for classifying different road semantics discovered from {\bfseries in-vehicle} traces.}
\centering
\small
\begin{tabular}{||c||c|c|c|c|c|c|c||c||c|c||c||} \hline
\hline
	& Cat's eyes	& Bumps	&Curves	&Rail cross. &Bridges	&Tunnels & Tur.\&cur.&unclass.&FP	&FN	&$\sum$\\\hline\hline
Cat's eyes	& \cellcolor{gray!10}{\bfseries 22}&0&0&0&0&0&0&5&0&0.18	& 27\\\hline
Bumps	& 0&\cellcolor{gray!10}{\bfseries 30}&0&3 &0&	0&0&0&0.03 &	0.09 &	33\\\hline
Curves	&0&0&\cellcolor{gray!10}{\bfseries 20}	&0&0&0&0&0&0&0&20\\\hline
Rail cross. &	0&1	&0&\cellcolor{gray!10}{\bfseries 13}	& 0&0&0&0	&0.21&0.07 &	14\\\hline
Bridges	&0&0&0&0&\cellcolor{gray!10}{\bfseries 9	}&0&0&1&0&0.1	&10\\\hline
Tunnels	&0&0&0&0&0&\cellcolor{gray!10}{\bfseries 11}&0&0&0&0&11\\\hline
Tur.\&cur. (Rndabt )	&0&0&0&0&0&0&\cellcolor{gray!10}{\bfseries 41}&0&0&0&41\\\hline\hline
Overall	&\multicolumn{8}{l}{}				 & \cellcolor{gray!20}{\bfseries 0.03}	& \cellcolor{gray!20}{\bfseries 0.06}	&\cellcolor{gray!20}{\bfseries 156}\\\hline
\hline\end{tabular}
\label{tab:drive_conf}
\end{table*}

\begin{table*}[!t]
\caption{Confusion matrix for classifying different road semantics discovered from {\bfseries pedestrian} traces.}
\centering
\small
\begin{tabular}{||c||c|c|c|c|c|c|c||c|c||c||} \hline
\hline
	&Underpass& Stairs&Escalator&Footbridge&Walking&Stationary&Crosswalk&FP	&FN	&$\sum$\\\hline\hline
Underpass& \cellcolor{yellow!25}{\bfseries 11}&0&0&0&0&0&0&0&0	& 11\\\hline
Stairs	& 0&\cellcolor{yellow!25}{\bfseries 14}&0&0&2&0&0&0 &	0.13 &	16\\\hline
Escalator	&0&0&\cellcolor{yellow!25}{\bfseries 15}	&0&0&0&0&0&0&15\\\hline
Footbridge &	0&0	&0&\cellcolor{yellow!25}{\bfseries 16}	&0&0& 1	&0&0.06 &	17\\\hline
Walking	&0&0&0&0&\cellcolor{yellow!25}{\bfseries 32 }&0&0&0.06	&0&32\\\hline
Stationary	&0&0&0&0&0&\cellcolor{yellow!25}{\bfseries 15}&0&0&0&15\\\hline
Crosswalk	&0&0&0&0&0&0&\cellcolor{yellow!25}{\bfseries 10 }&0.1&0	&10\\\hline
Overall	&\multicolumn{7}{l}{}				 & \cellcolor{yellow!35}{\bfseries 0.02}	& \cellcolor{yellow!35}{\bfseries 0.03}	&\cellcolor{yellow!35}{\bfseries 116}\\\hline
\hline\end{tabular}
\label{tab:ped_conf}
\end{table*}
\subsection{Discovered Semantic Road Features Location Accuracy}
\begin{figure}[!t]
\centering
\includegraphics[width=0.65\linewidth]{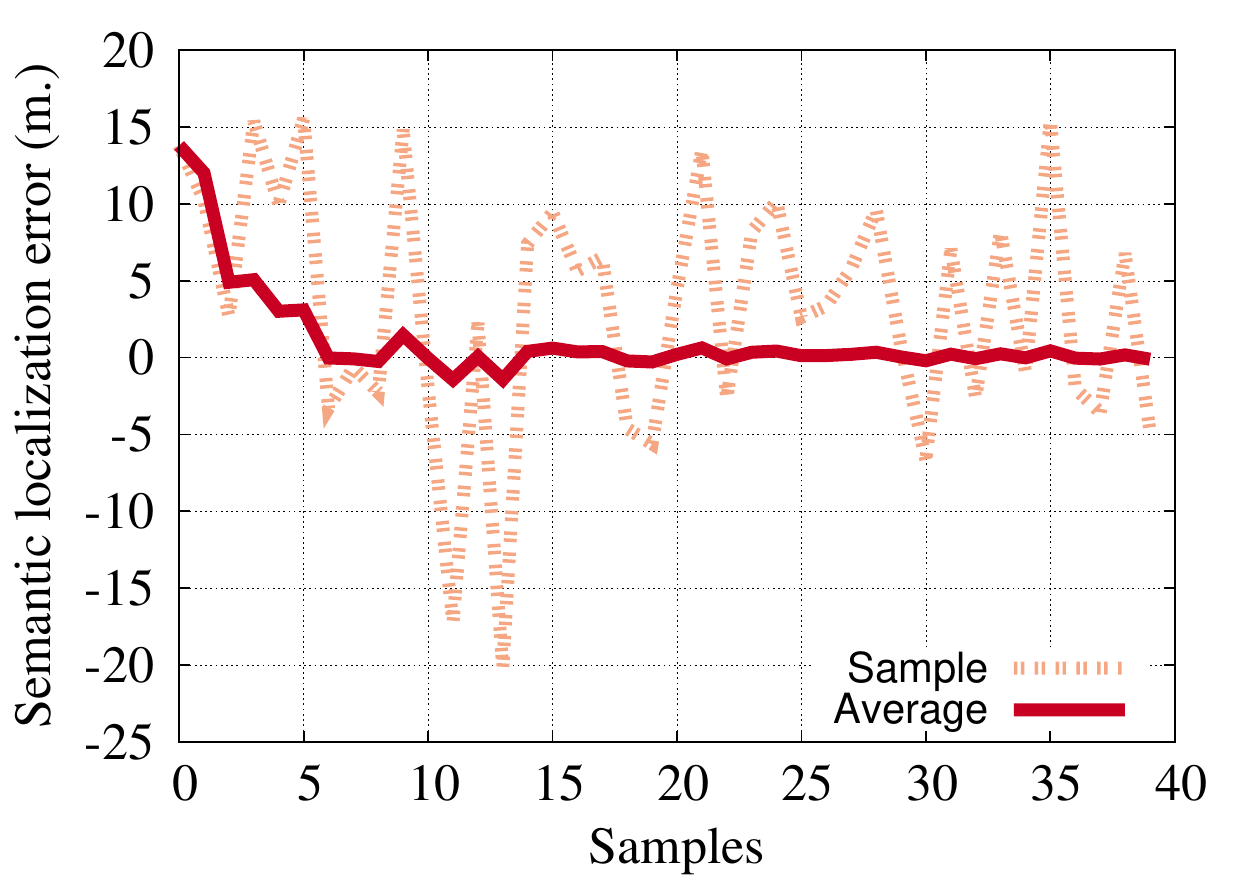}
\caption{Effect of number of samples on the accuracy of estimating the semantic location.}
\label{fig:veh_anchor_loc}
\end{figure}

Figure \ref{fig:veh_anchor_loc} shows that the errors in the location of the discovered map feature drop quickly as we increase the number of crowd-sensed samples. We can consistently reach an accuracy of less than 2~m using as few as 15 samples for all discovered map features.

\begin{figure}[!t]
\centering
\includegraphics[width=0.65\linewidth]{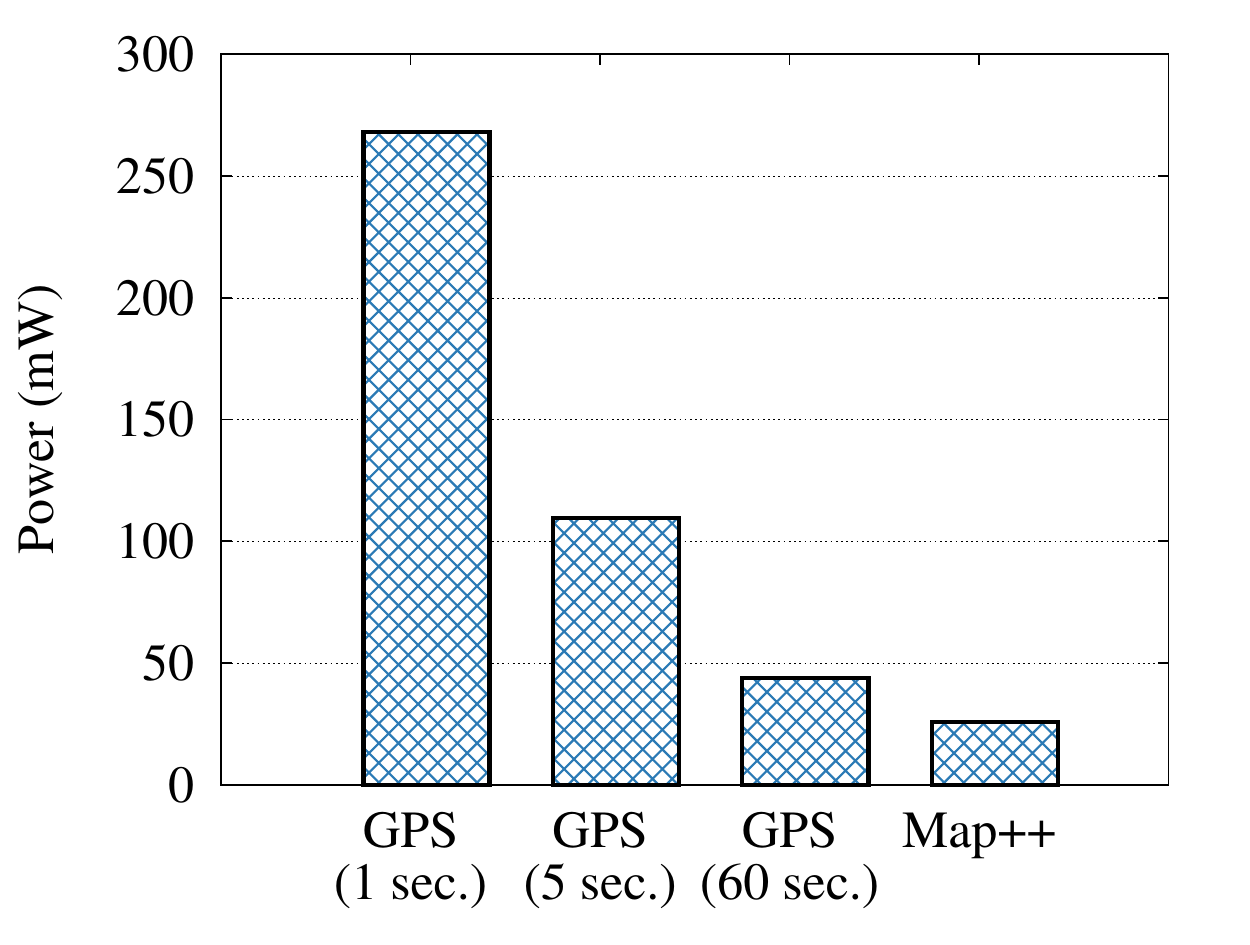}
\caption{Energy footprint of \system{} as compared to systems that use the GPS\cite{crowdAtlas,cao2009gps,baier2011mapcorrect} with different duty cycles.}
\label{fig:power}
\end{figure}

\subsection{Power Consumption}
Figure \ref{fig:power} shows the power consumption of \system{}, which is based on the inertial sensors for both road semantic detection and localization based on \emph{Dejavu}, as compared to systems that detects the missing road segments only, based on the GPS traces \cite{crowdAtlas,cao2009gps,baier2011mapcorrect} with different duty-cycles. The power is calculated using the PowerTutor profiler \cite{powertutor} and the android APIs using the HTC Nexus One cell phone. The figure shows that \system{} has a significantly lower energy profile compared to systems that are based on the GPS chip. Note that since inertial sensors are used during the normal phone operation, to detect the phone orientation change, \system{} practically consumes zero extra sensing power in addition to the standard phone operation.

\begin{table*}[!t]
\centering
\small
\caption{Summary for state-of-the-art digital map update techniques.}
\begin{tabular}{|p{3.4cm}||p{2.5cm}|p{8.5cm}|p{2.5cm}|}
\hline
\backslashbox{Technique}{Criteria} & Sensors Used & Features Added & Consumed Power\\\hline\hline
CrowdAtlas\cite{crowdAtlas} & GPS traces & New roads, road direction, and road type. & GPS power \\\hline
SmartRoad\cite{hu2013smartroad} & GPS traces & Unregulated intersections, intersections with traffic lights, and intersections with stop signs. & GPS power \\\hline
Carisi et al. \cite{carisi2011enhancing} & GPS traces & Traffic lights and stop signs. & GPS power \\\hline\hline
{\bfseries \system{} }& {\bfseries Inertial Sensors, GSM, Dejavu localization }&{\bfseries Tunnels, bridges, traffic calming, railway crossing, stop signs, underpasses,  footbridges, stairs, bridge height, crosswalk, roundabouts, and number of lanes.} &{\bfseries Inertial sensors power}\\\hline
\end{tabular}
\label{tab:rw_comp}
\end{table*}

\section{Related Work}\label{sec:rel_work}
Table~\ref{tab:rw_comp} compares \system{} to the state-of-the-art digital map update techniques. 

\subsection{Digital Map Update}
Digital maps contain inaccuracies that may limit their value to different applications. Automatic map update is a promising solution for these inaccuracies. Recent work introduced map inference using user trajectories for indoor and outdoor digital maps. For outdoor maps, raw GPS traces were used to update the maps. For example, in CrowdAtlas\cite{crowdAtlas}, authors used a map matching approach to infer missing roads on OSM using raw GPS traces. They also identified turns locations on the map. Similarly, in \cite{carisi2011enhancing}, authors proposed a system to extend digital maps with the location and timing of stop-signs and traffic lights in a city, using GPS traces collected by vehicles. \system{} complements these systems and uses a larger set of sensors, allowing it to detect significantly more semantic features. In addition, the sensors it uses for features detection and localization have a significantly lower-energy profile than GPS. 

For indoor maps, recently in \cite{youssef2012ubiquitous,alzantot2012crowdinside,alzantot2013demonstrating} authors proposed indoor maps construction and inferring indoor structures like elevators using sensors available on smart-phones. \system{} uses a similar approach for semantics inference. However, outdoor maps have completely different diverse semantics.

\subsection{Road Semantics}
Recently, inertial sensors embedded in smart phones allowed detection of the different road features. 
In\cite{aly2013dejavu}, we showed that a \emph{finite state machine} could be applied to inertial sensors to recognize different physical and logical landmarks (e.g. bridges, turns, and cellular signal anomalies) for \emph{in-vehicle} mobile phones. The goal was to provide an accurate and energy-efficient GPS replacement. \system{} extends this work by adding more semantic features, such as roundabouts, as well as using a classifier-based approach, which provides the \emph{same accuracy} in a more intuitive way with simpler implementation and more compact representation. In addition, \system{} enriches the road semantic by a novel class of pedestrian-based semantic features such as underpasses, footbridges, road capacity among others.

Monitoring road condition using inertial sensors was proposed in \cite{pothole,nericell}.
They mainly use the inertial sensors to detect the potholes and traffic conditions and use GPS to localize
the sensed road problems.  Both systems use external sensor chips which have higher sampling rates and lower noise compared to chips on typical cell-phones in the market. In addition, they depend on the energy-hungry GPS. \system, on the other hand, detects a significantly richer set of features, both based on vehicle and pedestrian traces, using a lower energy-profile sensors available in commodity cell-phones.

\section{Conclusion}\label{sec:conc}
In this paper we presented \system{}: a system for automatically enriching digital maps via a crowdsensing approach based on standard cell phones. For energy efficiency,  \system{} uses only low-energy sensors and sensors that are already running for other purposes. We presented the \system{} architecture as well as the features and classifiers that can accurately detect the different road features such as tunnels, bridges, crosswalks, stairs, and footbridges from the user traces.  

We implemented our system using commodity mobile phones running the Android operating system and evaluated it in the city of Alexandria, Egypt. Our results show that we can detect the map semantics accurately with 3\% false positive rate and 6\% false negative for in-vehicle traces and  2\% false positive rate and 3\% false negative for pedestrian traces. In addition, \system{} has a significantly lower energy profile compared to systems that are based on GPS.

Currently, we are expanding the system in multiple directions including
inferring higher level semantic information, such as different point of interests (POIs), using more available sensors and optimizing their power consumption, inferring more road features, among others.
\section{Acknowledgment}
This work was supported in part by the KACST National Science and Technology Plan under grant \#11-INF2062-10, and the KACST GIS Technology Innovation Center at Umm Al-Qura University under grant \#GISTIC-13-09.

\bibliographystyle{abbrv}
\tiny
\bibliography{map++}

\begin{thebibliography}{10}

\bibitem{bingmaps}
Bing {M}aps.
\newblock http://www.bing.com/maps/.

\bibitem{ustrsnp}
Federal {H}ighway {A}dministration ({FHWA}).
\newblock http://www.fhwa.dot.gov/.

\bibitem{googleio}
Google {I/O} 2013 session ({G}oogle {M}aps: {I}nto the future).
\newblock https://www.youtube.com/watch?v=sBAd89C4Q8Q.

\bibitem{googlemapmaker}
Google {M}ap {M}aker.
\newblock http://www.google.com/mapmaker.

\bibitem{googlemaps}
Google {M}aps.
\newblock http://maps.google.com/.

\bibitem{heremapcreator}
Here {M}ap {C}reator.
\newblock http://here.com/mapcreator/.

\bibitem{navteqhist}
{NAVTEQ} history.
\newblock http://corporate.navteq.com/company\_history.htm.

\bibitem{osm}
Open{S}treet{M}ap.
\newblock http://www.openstreetmap.org/.

\bibitem{wikimapia}
Wikimapia.
\newblock http://www.wikimapia.org.

\bibitem{yahoomaps}
Yahoo! {M}aps.
\newblock http://maps.yahoo.com/.

\bibitem{aly2013dejavu}
H.~Aly and M.~Youssef.
\newblock Dejavu: An accurate energy-efficient outdoor localization system.
\newblock In {\em SIGSPATIAL}. ACM, 2013.

\bibitem{alzantot2012crowdinside}
M.~Alzantot and M.~Youssef.
\newblock Crowd{I}nside: automatic construction of indoor floorplans.
\newblock In {\em SIGSPATIAL}. ACM, 2012.

\bibitem{alzantot2012uptime}
M.~Alzantot and M.~Youssef.
\newblock Uptime: {U}biquitous pedestrian tracking using mobile phones.
\newblock In {\em WCNC}. IEEE, 2012.

\bibitem{alzantot2013demonstrating}
M.~Alzantot and M.~Youssef.
\newblock Demonstrating {C}rowd{I}nside: {A} system for the automatic
  construction of indoor floor-plans.
\newblock In {\em PERCOM Workshops}. IEEE, 2013.

\bibitem{baier2011mapcorrect}
P.~Baier, H.~Weinschrott, F.~Durr, and K.~Rothermel.
\newblock Map{C}orrect: {A}utomatic correction and validation of road maps
  using public sensing.
\newblock In {\em Local Computer Networks (LCN)}. IEEE, 2011.

\bibitem{cao2009gps}
L.~Cao and J.~Krumm.
\newblock From {GPS} traces to a routable road map.
\newblock In {\em SIGSPATIAL}. ACM, 2009.

\bibitem{carisi2011enhancing}
R.~Carisi, E.~Giordano, G.~Pau, and M.~Gerla.
\newblock Enhancing in vehicle digital maps via {GPS} crowdsourcing.
\newblock In {\em Wireless On-Demand Network Systems and Services (WONS)}.
  IEEE, 2011.

\bibitem{wifiacc}
Y.-C. Cheng, Y.~Chawathe, A.~LaMarca, and J.~Krumm.
\newblock Accuracy characterization for metropolitan-scale {Wi-Fi}
  localization.
\newblock In {\em MobiSys}, pages 233--245. ACM, 2005.

\bibitem{cleveland1988locally}
W.~S. Cleveland and S.~J. Devlin.
\newblock Locally weighted regression: {A}n approach to regression analysis by
  local fitting.
\newblock {\em Journal of the American Statistical Association}, 83, 1988.

\bibitem{pothole}
J.~Eriksson, L.~Girod, B.~Hull, R.~Newton, S.~Madden, and H.~Balakrishnan.
\newblock The pothole patrol: {U}sing a mobile sensor network for road surface
  monitoring.
\newblock In {\em MobiSys}. ACM, 2008.

\bibitem{ester1996density}
M.~Ester, H.-P. Kriegel, J.~Sander, and X.~Xu.
\newblock A density-based algorithm for discovering clusters in large spatial
  databases with noise.
\newblock In {\em KDD}, 1996.

\bibitem{hu2013smartroad}
S.~Hu, L.~Su, H.~Liu, H.~Wang, and T.~Abdelzaher.
\newblock Smart{R}oad: {A} crowd-sourced traffic regulator detection and
  identification system.
\newblock In {\em Proc. of the int. conf. on Inf. processing in sensor
  networks}. ACM, 2013.

\bibitem{cellsense2}
M.~Ibrahim and M.~Youssef.
\newblock Cell{S}ense: {A} probabilistic {RSSI}-based {GSM} positioning system.
\newblock In {\em GLOBECOM}, pages 1--5. IEEE, 2010.

\bibitem{ibrahim2011hidden}
M.~Ibrahim and M.~Youssef.
\newblock A hidden markov model for localization using low-end {GSM} cell
  phones.
\newblock In {\em ICC}. IEEE, 2011.

\bibitem{cellsense}
M.~Ibrahim and M.~Youssef.
\newblock Cell{S}ense: {A}n accurate energy-efficient {GSM} positioning system.
\newblock {\em IEEE T. Vehicular Technology}, 2012.

\bibitem{ibrahim2013enabling}
M.~Ibrahim and M.~Youssef.
\newblock Enabling wide deployment of {GSM} localization over heterogeneous
  phones.
\newblock In {\em ICC}. IEEE, 2013.

\bibitem{krumm2004tempio}
J.~Krumm and R.~Hariharan.
\newblock Tempio: {I}nside/outside classification with temperature.
\newblock In {\em 2nd Int. Workshop on Man-Machine Symbiotic sys.}, 2004.

\bibitem{powertutor}
Z.~M. L.~Zhang, R.~Dick and L.~Yang.
\newblock Power{T}utor: {A} power monitor for android-based mobile platforms.
\newblock [Online]. Available: http://powertutor.org.

\bibitem{nericell}
P.~Mohan, V.~N. Padmanabhan, and R.~Ramjee.
\newblock Nericell: {R}ich monitoring of road and traffic conditions using
  mobile smartphones.
\newblock In {\em SenSys}. ACM, 2008.

\bibitem{ravindranath2011improving}
L.~Ravindranath, C.~Newport, H.~Balakrishnan, and S.~Madden.
\newblock Improving wireless network performance using sensor hints.
\newblock In {\em Proc. of the USENIX conf. on Networked sys. design and
  implementation}, 2011.

\bibitem{trnsp_mode_1}
S.~Reddy, M.~Mun, J.~Burke, D.~Estrin, M.~Hansen, and M.~Srivastava.
\newblock Using mobile phones to determine transportation modes.
\newblock {\em ACM Transactions on Sensor Networks (TOSN)}, 6(2), 2010.

\bibitem{trnsp_mode_3}
L.~Stenneth, O.~Wolfson, P.~S. Yu, and B.~Xu.
\newblock Transportation mode detection using mobile phones and {GIS}
  information.
\newblock In {\em SIGSPATIAL}. ACM, 2011.

\bibitem{tang2012efficient}
Y.~Tang, A.~D. Zhu, and X.~Xiao.
\newblock An efficient algorithm for mapping vehicle trajectories onto road
  networks.
\newblock In {\em SIGSPATIAL}. ACM, 2012.

\bibitem{crowdAtlas}
Y.~Wang, X.~Liu, H.~Wei, G.~Forman, and Y.~Zhu.
\newblock Crowd{A}tlas: {S}elf-updating maps for cloud and personal use.
\newblock MobiSys '13.

\bibitem{youssef2012ubiquitous}
M.~Youssef, M.~Elzantout, R.~Elkhouly, and A.~Lotfy.
\newblock Ubiquitous indoor localization and worldwide automatic construction
  of floor plans.
\newblock {\em arXiv preprint arXiv:1204.3328}, 2012.

\bibitem{GAC}
M.~Youssef, M.~A. Yosef, and M.~El-Derini.
\newblock {GAC}: energy-efficient hybrid {GPS}-accelerometer-compass {GSM}
  localization.
\newblock In {\em GLOBECOM}. IEEE, 2010.

\bibitem{trnsp_mode_2}
Y.~Zheng, Q.~Li, Y.~Chen, X.~Xie, and W.-Y. Ma.
\newblock Understanding mobility based on {GPS} data.
\newblock Ubicomp. ACM, 2008.

\bibitem{zheng2008learning}
Y.~Zheng, L.~Liu, L.~Wang, and X.~Xie.
\newblock Learning transportation mode from raw {GPS} data for geographic
  applications on the web.
\newblock In {\em Proc. of the 17th int. conf. on World Wide Web}. ACM, 2008.

\bibitem{zhou2012iodetector}
P.~Zhou, Y.~Zheng, Z.~Li, M.~Li, and G.~Shen.
\newblock Iodetector: {A} generic service for indoor outdoor detection.
\newblock In {\em Proc. of the ACM Conf. on Embedded Network Sensor Systems},
  2012.

\end{thebibliography}
\end{document}